\documentclass[11pt,preprint]{aastex61}

\usepackage{natbib}
\usepackage {float}
\usepackage{graphicx}
\usepackage{amssymb,amsmath,mathtools}

\received{}
\accepted{}
\shorttitle{DR2 and the Hubble Constant}
\shortauthors{Riess et al.}

\newcommand{\xddots}{%
  \raise 5pt \hbox {.}
  \mkern 6mu
  \raise 1pt \hbox {.}
  \mkern 6mu
  \raise -3pt \hbox {.}
}

\newcommand{\kmsmpc}{\hbox{$ \, \rm km\, s^{-1} \, Mpc^{-1}$}}

\newcommand{\bq}{\begin{equation}} 
\newcommand{\eq}{\end{equation}}

\newcommand{\diffv}{0.024 mag, SD=0.032 mag}
\newcommand{\diffi}{0.038 mag, SD=0.027 mag}
\newcommand{\diffh}{-0.056 mag, SD=0.048 mag}
\newcommand{\diffw}{-0.051 mag, SD=0.052 mag}

\newcommand{\facezp}{$-46 \pm 13 \,\mu$as \,}
\newcommand{\facemult}{$1.006 \pm 0.033 \,$}
\newcommand{\facechi}{62.5}
\newcommand{\faceplanck}{2.9}

\newcommand{\noscalezp}{$ -46 \pm 6 \,\mu$as}

\newcommand{\expandzp}{$-47 \pm 16 \,\mu$as}
\newcommand{\expandmult}{$1.008 \pm 0.039$}
\newcommand{\expandchi}{45.0}
\newcommand{\expandplanck}{2.6}

\newcommand{\beq}{\begin{equation}}
\newcommand{\eeq}{\end{equation}}
\newcommand{\beqa}{\begin{eqnarray}}
\newcommand{\eeqa}{\end{eqnarray}}

\newcommand{\PL}{$P\hbox{--}L$}

\long\def\check#1{}
\long\def\hide#1{}

\newcommand{\Gaia}{{\it Gaia}}

\begin{document} 

\title{Milky Way Cepheid Standards for Measuring Cosmic Distances and Application to {\Gaia} DR2: \\ Implications for the Hubble Constant}

\author{Adam G.~Riess}
\affiliation{Space Telescope Science Institute, 3700 San Martin Drive, Baltimore, MD 21218, USA}
\affiliation{Department of Physics and Astronomy, Johns Hopkins University, Baltimore, MD 21218, USA}

\author{Stefano Casertano}
\affiliation{Space Telescope Science Institute, 3700 San Martin Drive, Baltimore, MD 21218, USA}
\affiliation{Department of Physics and Astronomy, Johns Hopkins University, Baltimore, MD 21218, USA}

\author{Wenlong Yuan}
\affiliation{Department of Physics and Astronomy, Johns Hopkins University, Baltimore, MD 21218, USA}
\affiliation{Texas A\&M University, Department of Physics and Astronomy, College Station, TX, USA}

\author{Lucas Macri}
\affiliation{Texas A\&M University, Department of Physics and Astronomy, College Station, TX, USA}

\author{Beatrice Bucciarelli}
\affiliation{INAF, Osservatorio Astronomico di Torino, Strada Osservatorio 20, 
     I-10025 Pino Torinese, TO, Italy}
     
\author{Mario G.~Lattanzi}
\affiliation{INAF, Osservatorio Astronomico di Torino, Strada Osservatorio 20, 
     I-10025 Pino Torinese, TO, Italy}

\author{John W.~MacKenty}
\affiliation{Space Telescope Science Institute, 3700 San Martin Drive, Baltimore, MD 21218, USA}

\author{J.~Bradley Bowers}
\affiliation{Department of Physics and Astronomy, Johns Hopkins University, Baltimore, MD 21218, USA}
\affiliation{Department of Physics and Astronomy, Johns Hopkins University, Baltimore, MD 21218, USA}

\author{WeiKang Zheng}
\affiliation{Department of Astronomy, University of California, Berkeley, CA 94720-3411, USA}

\author{Alexei V.~Filippenko}
\affiliation{Department of Astronomy, University of California, Berkeley, CA 94720-3411, USA}
\affiliation{Miller Senior Fellow, Miller Institute for Basic Research in Science, University of California, Berkeley, USA}

\author{Caroline Huang}
\affiliation{Department of Physics and Astronomy, Johns Hopkins University, Baltimore, MD 21218, USA}

\author{Richard I.~Anderson}
\affiliation{European Southern Observatory, Karl-Schwarzschild-Str. 2, 85748 Garching b. M\"unchen, Germany}

\begin{abstract} 

We present {\it Hubble Space Telescope (HST)} photometry of a selected sample of 50 long-period, low-extinction Milky Way Cepheids measured on the same WFC3 $F555W$-, $F814W$-, and $F160W$-band photometric system as extragalactic Cepheids in Type Ia supernova host galaxies.  These bright Cepheids were observed with the WFC3 spatial scanning mode in the optical and near-infrared to mitigate saturation and reduce pixel-to-pixel calibration errors to reach a mean photometric error of 5 millimags per observation.  We use the new {\Gaia} DR2 parallaxes and {\it HST} photometry to simultaneously constrain the cosmic distance scale and to measure the DR2 parallax zeropoint offset appropriate for Cepheids.  We find the latter to be \facezp or $\pm$ 6 $\mu$as for a fixed distance scale, higher than found from quasars, as expected, for these brighter and redder sources.  The precision of the distance scale from DR2 has been reduced by a factor of 2.5 because of the need to independently determine the parallax offset. The best-fit distance scale is \facemult, relative to the scale from \cite{Riess:2016} with H$_0=73.24$ \kmsmpc\ used to predict the parallaxes photometrically, and is inconsistent with the scale needed to match the Planck 2016 CMB data combined with $\Lambda$CDM at the \faceplanck$\sigma$ confidence level (99.6\%).   At 96.5\% confidence we find that the formal DR2 errors may be underestimated as indicated. We identify additional error associated with the use of augmented Cepheid samples utilizing ground-based photometry and discuss their likely origins.  Including the DR2 parallaxes with all prior distance-ladder data raises the current tension between the late and early Universe route to the Hubble constant to 3.8$\sigma$ (99.99\%).  With the final expected precision from {\it Gaia}, the sample of 50 Cepheids with {\it HST} photometry will limit to 0.5\% the contribution of the first rung of the distance ladder to the uncertainty in the H$_0$.

\end{abstract} 


\section{Introduction} 

Measurements of cosmic distances from standard candles form a cornerstone of our cosmological model.   Yet even the best available standard candles require calibration of their absolute brightnesses via geometric distance measurements.  Trigonometric parallax is the ``gold standard" of such geometric distance measurements --- the simplest, the most direct, and the most assumption-free.  Previously, most Milky Way (MW) stars, including known examples of rare stars, were well out of range of even state-of-the-art 0.3--1 milliarcsecond (mas) parallax measurements from {\it Hipparcos} and the Fine Guidance Sensor (FGS) aboard the {\it Hubble Space Telescope\/} ({\it HST\/}).  Now, we are entering a ``golden age" of parallax determinations as most of the MW's stars will come into parallax range from relative astrometry measured at the $\mu$as level by the ESA mission, {\Gaia} \citep{Gaia-Collaboration:2016,Gaia-Collaboration:2016a,Gaia-Collaboration:2018}.  

The parallaxes of long-period Cepheids are among the most coveted because these variables can be seen with {\it HST} in the host galaxies of Type Ia supernovae (SNe~Ia) at $D< 50$ Mpc and used to calibrate their luminosities and the expansion rate of the Universe \citep{Riess:2016} (hereafter R16).  \cite{benedict07} used the FGS on {\it HST} to measure parallaxes to 9 of the 10 Cepheids in the MW known at $D\!<\!0.5$ kpc with individual precision of 8\% and a sample mean error of 2.5\%.  However, all but one of these had periods $<10$ days, a range where Cepheids are too faint to be observed in most SN Ia hosts.  Nearly all of the long-period Cepheids live at $D>1$ kpc, demanding a parallax precision better than $ 100 \,\mu$as for a useful measurement.      Spatial scanning with {\it HST}'s WFC3 has provided relative astrometry with 30--40 $\mu$as precision to extend the useful range of Cepheid parallaxes to 2--4 kpc, measuring 8 with $P \geq 10$ days with an error in the mean distance of 3\% and providing a calibration more applicable to extragalactic Cepheid samples \citep{riess14, Casertano:2016, Riess:2018}.  
      
The {\Gaia} mission is expected to measure the parallaxes of hundreds of MW Cepheids with a precision of 5--$10\,\mu$as by the end of the mission.  Such parallax measurements would support a $\sim 1\% $ determination of the Hubble constant (H$_0$) {\it provided the possible precision of the calibration of Cepheid luminosities is not squandered by photometric inaccuracy}. To retain the precision of {\Gaia}'s Cepheid parallaxes when they are used as standard candles it is necessary to measure their mean brightness on the same photometric systems used to measure their extragalactic counterparts.  By using such purely {\it differential} flux measurements of Cepheids along the distance ladder, it is possible to circumvent systematic uncertainties related to zeropoints and transmission functions which otherwise incur a {\it systematic} uncertainty of $\sim$ 2--3\% in the determination of H$_0$, nearly twice the target goal, even before including additional uncertainties along the distance ladder. 

To forge this photometric bridge, in {\it HST} Cycle 20 (2012) we began observing MW Cepheid ``Standards'' among the set of 70  with $P>$ 8 days, $A_H < 0.4 $ mag, $V > 6$ mag and expected distances of $D < 7$ kpc, criteria which yield the most useful sample for calibration of the thousands of extragalactic Cepheids observed in the hosts of SNe~Ia.  These extragalactic Cepheids across the hosts of 19 SNe~Ia and in NGC 4258 have all been observed in the near-infrared (NIR) with WFC3-IR in filter $F160W$ (similar to the $H$ band) to reduce systematics caused by reddening and metallicity, and in optical colors $F555W$ (similar to the $V$ band) and $F814W$ (similar to the $I$ band), to form a reddening-free distance measure \citep{Hoffmann:2016,Riess:2016}.  To measure the much closer and brighter MW Cepheids on the same photometric system and mitigate saturation, we used very fast spatial scans, moving the telescope during the observation so that the target covers a long, nearly vertical line over the detector.  We use a scan speed of $ 7\farcs 5 / {\rm sec} $, corresponding to an effective exposure time of $ 0.005 \, \rm sec $ in the visible and $ 0.02 \,\rm sec $ in the infrared, much shorter than the minimum effective exposure times possible with the WFC3 hardware.  Scanning observations are also free from the variations and uncertainties in shutter flight time (for $F555W$ and $F814W$ with WFC3-UVIS) that affect very short pointed observations \citep{Sahu:2015}.  Spatial scans offer the additional advantage of varying the position of the source on the detector, which averages down pixel-to-pixel errors in the flat fields, and can also be used to vary the pixel phase, reducing the uncertainty from undersampled point-spread-function photometry.  Finally, unlike ground-based photometry which relies on calibrators in the same region of the sky, {\it HST} can measure the photometry of MW Cepheids over the whole sky, without concern about regional variations in calibrators.  Most of the observations were obtained in {\it HST} ``SNAP'' mode, which selects a subset of the targets to be observed based on scheduling convenience, thus essentially randomly with respect to intrinsic Cepheid properties.  Observations were obtained for a total of 50 of the 70 Cepheids, which therefore constitute an unbiased subset of the full sample.

In \S~2 we present the 3-filter spatial scan photometry of the 50 MW Cepheids observed in our {\it HST} programs, and compare them internally as well as with ground-based measurements in corresponding passbands.  In \S~3 we carry out an analysis of the recently released {\Gaia} DR2 parallaxes for our targets; using the precise and accurate {\it HST} photometry, we verify the existence and magnitude of a zeropoint offset for the {\Gaia} parallaxes, and at the same time test current measurements of H$_0$.  In \S~4 we discuss these results and the nature of the zeropoint issue.

\section{Milky Way Cepheid Standards}

   The 50 MW Cepheids collected here were observed photometrically in several {\it HST} programs: GO-12879, GO-13334, GO-13686, GO-13678, GO-14206, and GO-14268 include photometric and astrometric measurements for 18 of the targets, while GO-13335 and GO-13928 were purely photometric SNAP programs.  For 8 of these targets, the photometric measurements have been reported in \cite{Riess:2018}; the photometry of the other 42 targets follows the same procedures.  Here we summarize the key steps for convenience; the full description is in \cite{Riess:2018}. 

1. Fluxes are measured from the amplitude of the fits of the line-spread function to the extracted signal at every position along the scan; a 15-pixel minirow across the scan is used to perform the fit.  The flux is divided by the effective exposure time, i.e., the pixel size divided by the scan rate.  Pairs of direct and scanning mode images are used to calibrate out possible errors in the pixel size and scan rate, and provide the aperture correction applicable between scanning and staring mode observations.  This offset has an error in the mean of 0.002--0.003 mag, depending on the filter.

2. We multiply the measured flux by the local (relative) pixel area using the same pixel area map used for photometry of all point sources in staring mode; this corrects from flux per unit area to actual flux.
   
3. We then need to correct for the differing sizes of the pixel length along the scan ($Y$) direction, which changes the effective exposure time seen at each location along the scan.  This step partially reverses the correction in step 2; the net result of steps 2 and 3 is to multiply the fitted amplitude by only the relative pixel size perpendicular to the scan direction.  \cite{Riess:2018} compare pairs of scans of MW Cepheids in back-to-back exposures, and demonstrate a mean photometric error per scan observation of 0.007, 0.003, and 0.001 mag in $F160W$, $F555W$, and $F814W$, respectively.  For the sample of 50 Cepheids presented here, the mean number of epochs per filter is between 2 and 3.  
  
4. Finally, we apply a correction for the light curve phase, i.e., the difference between each Cepheid's magnitude at the observed phase and the magnitude at the epoch of mean intensity of its light curve.  These phase corrections are derived from ground-based light curves of these Cepheids in filters with wavelengths best corresponding to the WFC3 filters with their sources given in a table in the Appendix. The phase corrections are calculated in the {\it HST} system after the ground-based light curves are transformed to this system using the transformations given in \cite{Riess:2016}. Because the phase corrections are relative quantities, they do not have a zeropoint and they do not change the zeropoint of the light curves, which remain on the {\it HST} WFC3 natural system. The uncertainties in these phase corrections depend on the quality of the ground-based light curves; the average uncertainty in the magnitude corrections is 0.024, 0.020, and 0.018 mag per epoch in $F555W$, $F814W$, and $F160W$, respectively.   The empirical scatter between multiple measurements for the same target --- typically 4--5 for the targets in \cite{Riess:2018} --- is consistent with the estimated uncertainties.  The mean uncertainty in the light-curve mean magnitude for these 50 Cepheids is 0.021, 0.018, and 0.015 mag in $F555W$, $F814W$, and $F160W$, respectively. The internal agreement between individual epochs, each corrected to the mean (for Cepheids with 3 or more epochs) is shown in Figure 1 and includes both the photometry errors and the phase-correction uncertainties.
  
For distance measurements and for the determination of H$_0$, it is useful to convert these three bands to the reddening-free Wesenheit magnitudes \citep{madore82} used by R16 for measuring extragalactic Cepheids in the hosts of Type Ia supernovae: 
\bq m^W_H=m_{F160W}-0.386(m_{F555W}-m_{F814W}). \eq
\noindent  These 50 $m^W_H$ values have a mean uncertainty of 0.019 mag, including photometric measurement errors, phase corrections, and error propagation to the Wesenheit magnitude, corresponding to approximately 1\% in distance; at the mean expected parallax of 400\ $\mu$as this represents a mean uncertainty of $ 4 \,\mu$as in the {\it predicted} parallax.  At this level of precision, both the breadth of the instability strip at 0.04--0.08 mag in $m^W_H$ \citep{Macri:2015,persson04} and the expected parallax uncertainties by the end of the {\Gaia} mission (5--$14\,\mu$as) will still dominate the determination of individual Cepheid luminosities.  Some of these Cepheids have been suggested as possible binaries, but in general we do not automatically exclude possible binaries from consideration. At a typical Milky Way Cepheid distance of 2.5 kpc,  companion separations of less than 0.1" for {\it HST} WFC3 UVIS channel or $<$ 400 AU are unresolved and thus included with the measured Cepheid flux.  This contribution, while small, is statistically matched in extragalactic Cepheids and thus cancels in the use of Cepheid fluxes along the distance ladder.  For wider binaries, $400-4000$ AU \cite{Anderson:2017} estimate that the effect on the photometric calibration of Cepheids is on the order of 0.004\% (in distance) and thus negligible.
    
In Table 1 we provide the photometric measurements of these 50 Cepheids for WFC3 $F555W$, $F814W$, $F160W$ and $m^W_H$.\footnote{For $m^W_H$ we include the correction for the count-rate nonlinearity effect (hereafter CRNL) for WFC3-IR to account for the 6.4 dex flux ratio in $F160W$ between these MW Cepheids and the sky-dominated extragalactic Cepheids \citep{Riess:2018}.}

The best-fit solution from \citet{Riess:2016} yields a calibration of the Cepheid $P$--$L$ relation of

\bq M^W_H=-5.93-3.26({\rm log} P - 1).  \eq 
\noindent
Employing the derived periods in Table 1 yields the values of $M^W_H$, and combined with the apparent Wesenheit magnitudes on the WFC3 system ($m^W_H$) we derive distance moduli of \bq \mu=m^W_H - M^W_H \eq
\noindent
and the expected parallax
\bq \pi_{R16}=10^{-0.2(\mu-10)} \eq in mas given in Table 1.      
With negligible uncertainties in the periods, the mean uncertainties in the predicted parallaxes are $\sim 2-3$\% in distance due to the width of the instability strip.  These expected parallaxes are on the scale in which H$_0=73.24$ \kmsmpc\ as obtained in the same best-fit solution from \citet{Riess:2016}.

This set of photometry offers a number of distinct advantages over ground-based magnitudes.  By measuring all Cepheids along the distance ladder (and in both hemispheres) with a single, stable photometric system, {\it HST} WFC3, we can largely eliminate the propagation of zeropoint and bandpass uncertainties among Cepheid flux measurements.  This is especially important in the NIR where individual system zeropoints are typically based on only a handful of historical standards, {\it systematic} uncertainties are $\sim 0.02$--0.03 mag \citep{riess11c}, and the relative systematic differences between two systems can be expected to be $\sim 0.03$--0.04 mag.  To illustrate these differences, we compare the {\it HST} WFC3 system photometry with ground-based equivalents, transformed into the same system using the conversions of \cite{Riess:2016}.  Ground-based observations in $V,I,J,H$ were obtained form the sources listed in the Appendix, and are rather inhomogeneous but the best available; the NIR measurements are primarily from three sources: \cite{Monson+2011}, \cite{Laney+1992}, and new observations obtained at CTIO.  In Figure 2 we compare the ground-based mean magnitudes to the {\it HST} WFC3 values.  We find mean differences (in the direction Ground$-$ {\it HST\/}) and a sample dispersion (SD) in $F555W$, $F814W$, $F160W$, and $m^W_H$ of \diffv, \diffi, \diffh, and \diffw, respectively; a few outliers (4, 2, 1, and 1, respectively) are marked in Figure~2.  In the following we use only the reddening-free Wesenheit magnitude $m^W_H$.

Restricting our analysis to {\it HST} magnitudes limits the sample of usable Cepheids; over 200 more have ground-based photometry, and in principle could be used for the same type of analysis, as was done for DR1 in \cite{Casertano:2017}.  However, our sample is close to complete for the most relevant Cepheids, those with long periods.  Moreover, at the much higher precision of DR2 vs. DR1 parallaxes (roughly 40 vs.~300 $ \mu$as per Cepheid), further reduced by averaging across a Cepheid sample, photometric errors and systematics become dominant, as we will demonstrate in \S~3.  Even at the current (DR2) precision, the quality of photometric information is paramount to obtain the best possible information from {\Gaia} parallaxes.

\startlongtable
\begin{deluxetable*}{ccccccccccccccc}
\tabletypesize{\scriptsize}
\tablewidth{0pt}
\tablenum{1}
\tablecaption{Photometric Data for MW Cepheids\label{tb:phot}}
\tablehead{\colhead{Cepheid} &  \colhead{log Period} & \colhead{$F555W$} & \colhead{$\sigma$}  & \colhead{$n$} & \colhead{$F814W$} & \colhead{$\sigma$}  & \colhead{$n$} & \colhead{$F160W^a$} & \colhead{$\sigma$}  & \colhead{$n$} & \colhead{$m^{W,b}_H$} & \colhead{$\sigma$}  & \colhead{$\pi_{R16}^{c}$} & \colhead{$\sigma$} \\
\colhead{} & \colhead{} & \colhead{(mag)} &  \colhead{} & \colhead{(epochs)} &  \colhead{(mag)} &  \colhead{} & \colhead{(epochs)} & \colhead{(mag)} &  \colhead{} & \colhead{(epochs)} &  \colhead{(mag)} &  \colhead{}  &  \colhead{(mas)}    &  \colhead{}}
\startdata
AA-GEM &  1.053  &  9.9130  &  0.029  &  1  &  8.542  &  0.025  &      1  &  7.348  &  0.017  &      1  &  6.871  &  0.023   &  0.254  &  0.005  \\
AD-PUP &  1.133  &  10.015  &  0.028  &  1  &  8.675  &  0.023  &      1  &  7.488  &  0.020  &      1  &  7.023  &  0.024   &  0.210  &  0.005  \\
AQ-CAR &  0.990  &  8.9836  &  0.020  &  2  &  7.854  &  0.009  &      2  &  6.766  &  0.007  &      3  &  6.382  &  0.011   &  0.350  &  0.007  \\
AQ-PUP &  1.479  &  8.8671  &  0.018  &  2  &  7.120  &  0.014  &      2  &  5.487  &  0.013  &      4  &  4.864  &  0.016   &  0.338  &  0.007  \\
BK-AUR &  0.903  &  9.5609  &  0.036  &  1  &  8.220  &  0.038  &      1  &  7.015  &  0.021  &      1  &  6.549  &  0.029   &  0.369  &  0.008  \\
BN-PUP &  1.136  &  10.051  &  0.033  &  1  &  8.505  &  0.017  &      1  &  7.198  &  0.015  &      1  &  6.653  &  0.021   &  0.248  &  0.005  \\
CD-CYG &  1.232  &  9.1207  &  0.011  &  3  &  7.468  &  0.012  &      3  &  5.900  &  0.012  &      5  &  5.314  &  0.014   &  0.398  &  0.008  \\
CP-CEP &  1.252  &  10.757  &  0.015  &  1  &  8.638  &  0.052  &      1  &  6.871  &  0.022  &      1  &  6.105  &  0.030   &  0.268  &  0.006  \\
CR-CAR &  0.989  &  11.750  &  0.019  &  1  &  9.973  &  0.018  &      1  &  8.384  &  0.014  &      1  &  7.750  &  0.017   &  0.187  &  0.004  \\
CY-AUR &  1.141  &  12.052  &  0.012  &  1  &  9.953  &  0.020  &      1  &  8.106  &  0.025  &      1  &  7.348  &  0.027   &  0.179  &  0.004  \\
DD-CAS &  0.992  &  10.036  &  0.007  &  3  &  8.523  &  0.011  &      3  &  7.108  &  0.012  &      4  &  6.576  &  0.013   &  0.319  &  0.006  \\
DL-CAS &  0.903  &  9.1059  &  0.019  &  1  &  7.569  &  0.022  &      1  &  6.238  &  0.018  &      1  &  5.697  &  0.021   &  0.547  &  0.011  \\
DR-VEL &  1.049  &  9.7083  &  0.034  &  1  &  7.770  &  0.020  &      1  &  6.183  &  0.021  &      1  &  5.487  &  0.026   &  0.484  &  0.011  \\
GQ-ORI &  0.935  &  8.7199  &  0.020  &  1  &  7.632  &  0.024  &      1  &  6.523  &  0.032  &      1  &  6.155  &  0.034   &  0.422  &  0.010  \\
HW-CAR &  0.964  &  9.2782  &  0.016  &  2  &  8.007  &  0.013  &      2  &  6.798  &  0.005  &      3  &  6.359  &  0.009   &  0.368  &  0.007  \\
KK-CEN &  1.086  &  11.598  &  0.017  &  1  &  9.862  &  0.021  &      1  &  8.292  &  0.015  &      1  &  7.674  &  0.018   &  0.167  &  0.003  \\
KN-CEN &  1.532  &  10.062  &  0.023  &  2  &  7.924  &  0.017  &      2  &  5.856  &  0.006  &      5  &  5.083  &  0.013   &  0.282  &  0.005  \\
RW-CAM$^*$ &  1.215  &  8.8673  &  0.015  &  1  &  7.044  &  0.014  &      1  &  5.451  &  0.021  &      1  &  4.799  &  0.022   &  0.517  &  0.011  \\
RW-CAS &  1.170  &  9.3719  &  0.021  &  1  &  7.863  &  0.016  &      1  &  6.483  &  0.022  &      1  &  5.952  &  0.024   &  0.326  &  0.007  \\
RY-CAS &  1.084  &  10.075  &  0.019  &  1  &  8.333  &  0.040  &      1  &  6.715  &  0.010  &      1  &  6.094  &  0.020   &  0.347  &  0.007  \\
RY-SCO &  1.308  &  8.2067  &  0.012  &  3  &  6.206  &  0.010  &      3  &  4.408  &  0.010  &      3  &  3.688  &  0.012   &  0.751  &  0.014  \\
RY-VEL$^*$ &  1.449  &  8.5234  &  0.036  &  1  &  6.757  &  0.016  &      1  &  5.211  &  0.017  &      1  &  4.581  &  0.023   &  0.403  &  0.009  \\
S-NOR &  0.989  &  6.5779  &  0.011  &  1  &  5.410  &  0.012  &      1  &  4.391  &  0.012  &      1  &  3.992  &  0.014   &  1.053  &  0.021  \\
S-VUL &  1.839  &  9.1668  &  0.008  &  3  &  6.862  &  0.012  &      3  &  4.885  &  0.010  &      4  &  4.047  &  0.011   &  0.287  &  0.006  \\
SS-CMA &  1.092  &  10.121  &  0.012  &  4  &  8.444  &  0.008  &      4  &  6.894  &  0.011  &      3  &  6.299  &  0.012   &  0.312  &  0.006  \\
SV-PER$^*$ &  1.046  &  9.2186  &  0.016  &  1  &  7.760  &  0.014  &      1  &  6.435  &  0.027  &      1  &  5.924  &  0.028   &  0.397  &  0.009  \\
SV-VEL &  1.149  &  8.7316  &  0.026  &  1  &  7.302  &  0.009  &      1  &  6.024  &  0.010  &      1  &  5.524  &  0.015   &  0.409  &  0.008  \\
SV-VUL &  1.653  &  7.2675  &  0.047  &  1  &  5.648  &  0.033  &      1  &  4.214  &  0.027  &      1  &  3.641  &  0.035   &  0.457  &  0.011  \\
SY-NOR &  1.102  &  9.8284  &  0.023  &  1  &  7.925  &  0.038  &      1  &  6.214  &  0.013  &      1  &  5.531  &  0.022   &  0.438  &  0.009  \\
SZ-CYG &  1.179  &  9.6209  &  0.013  &  2  &  7.756  &  0.017  &      2  &  6.004  &  0.008  &      3  &  5.336  &  0.012   &  0.427  &  0.008  \\
T-MON$^*$ &  1.432  &  6.0680  &  0.023  &  1  &  4.828  &  0.016  &      1  &  3.725  &  0.021  &      1  &  3.298  &  0.024   &  0.746  &  0.016  \\
U-CAR &  1.589  &  6.3852  &  0.038  &  1  &  4.967  &  0.023  &      1  &  3.768  &  0.019  &      1  &  3.272  &  0.026   &  0.596  &  0.013  \\
UU-MUS &  1.066  &  9.9212  &  0.024  &  1  &  8.457  &  0.025  &      1  &  7.108  &  0.010  &      1  &  6.595  &  0.017   &  0.283  &  0.006  \\
V0339-CEN &  0.976  &  8.8402  &  0.024  &  1  &  7.321  &  0.016  &      1  &  5.990  &  0.024  &      1  &  5.455  &  0.026   &  0.548  &  0.012  \\
V0340-ARA &  1.318  &  10.460  &  0.024  &  1  &  8.554  &  0.014  &      1  &  6.808  &  0.012  &      1  &  6.124  &  0.016   &  0.241  &  0.005  \\
VW-CEN &  1.177  &  10.379  &  0.031  &  1  &  8.718  &  0.023  &      1  &  7.158  &  0.010  &      1  &  6.569  &  0.018   &  0.243  &  0.005  \\
VX-PER &  1.037  &  9.4589  &  0.008  &  4  &  7.906  &  0.006  &      3  &  6.470  &  0.009  &      5  &  5.922  &  0.010   &  0.403  &  0.008  \\
VY-CAR &  1.276  &  7.6162  &  0.014  &  5  &  6.253  &  0.007  &      4  &  4.991  &  0.004  &      6  &  4.517  &  0.007   &  0.538  &  0.010  \\
VZ-PUP &  1.365  &  9.7715  &  0.033  &  1  &  8.262  &  0.022  &      1  &  6.931  &  0.017  &      1  &  6.400  &  0.023   &  0.198  &  0.004  \\
WX-PUP &  0.951  &  9.1909  &  0.030  &  1  &  7.944  &  0.012  &      1  &  6.807  &  0.010  &      1  &  6.378  &  0.016   &  0.372  &  0.007  \\
WZ-SGR &  1.339  &  8.2021  &  0.012  &  6  &  6.481  &  0.013  &      6  &  4.858  &  0.009  &      4  &  4.245  &  0.011   &  0.554  &  0.011  \\
X-CYG &  1.214  &  6.5295  &  0.020  &  1  &  5.230  &  0.049  &      1  &  4.080  &  0.033  &      1  &  3.630  &  0.039   &  0.887  &  0.023  \\
X-PUP &  1.414  &  8.6949  &  0.019  &  3  &  7.128  &  0.010  &      3  &  5.628  &  0.008  &      4  &  5.075  &  0.012   &  0.338  &  0.006  \\
XX-CAR &  1.196  &  9.4627  &  0.027  &  1  &  8.067  &  0.015  &      1  &  6.833  &  0.022  &      1  &  6.346  &  0.025   &  0.261  &  0.006  \\
XY-CAR &  1.095  &  9.4660  &  0.011  &  4  &  7.927  &  0.009  &      3  &  6.455  &  0.006  &      6  &  5.913  &  0.008   &  0.371  &  0.007  \\
XZ-CAR &  1.221  &  8.7725  &  0.017  &  3  &  7.217  &  0.006  &      3  &  5.770  &  0.007  &      4  &  5.221  &  0.010   &  0.422  &  0.008  \\
YZ-CAR &  1.259  &  8.8644  &  0.016  &  3  &  7.401  &  0.007  &      3  &  5.991  &  0.013  &      5  &  5.478  &  0.015   &  0.354  &  0.007  \\
YZ-SGR &  0.980  &  7.4662  &  0.021  &  1  &  6.176  &  0.014  &      1  &  5.103  &  0.020  &      1  &  4.657  &  0.022   &  0.786  &  0.017  \\
Z-LAC &  1.037  &  8.5686  &  0.022  &  1  &  7.157  &  0.015  &      1  &  5.917  &  0.018  &      1  &  5.424  &  0.021   &  0.507  &  0.011  \\
Z-SCT &  1.111  &  9.7535  &  0.019  &  2  &  8.079  &  0.021  &      2  &  6.513  &  0.008  &      3  &  5.918  &  0.014   &  0.362  &  0.007  \\
\hline
\enddata
\tablecomments{$^a$Does not include addition of $0.052 \pm 0.014$ mag to correct CRNL 6.4 dex between MW and extragalactic Cepheids.}
\tablecomments{$^b$Includes addition of $0.052 \pm 0.014$ mag to correct CRNL 6.4 dex between MW and extragalactic Cepheids.}
\tablecomments{$^c$ $\pi_{R16}=10^{-0.2(\mu-10)}$ where $\mu=m^W_H-M^W_H$ and $M^W_H$ is the absolute Wesenheit magnitude determined from the Cepheid period and the distance scale from \cite{Riess:2016} where $H_0$=73.24 \kmsmpc as discussed in the text.}
\tablecomments{$^*$ Not used in final analysis, see text}
\end{deluxetable*}

\section{{\Gaia} DR2}

The {\Gaia} mission \citep{prusti12, Gaia-Collaboration:2016, Gaia-Collaboration:2016a,Gaia-Collaboration:2018} is well positioned to revolutionize our knowledge of the luminosity scale of various stellar types, including those used to set the cosmic distance scale.  By mission end, {\Gaia} parallaxes for the Cepheids in our sample are expected to have errors of 5--14 $\mu$as, about 2\% of their typical parallax; with tens of objects, the uncertainty in the Cepheid luminosity calibration will be $ << 1\% $, negligible in the error budget for the local measurement of H$_0$ \citep{Riess:2016}.  

With the release of DR2 \citep[hereafter G18]{Gaia-Collaboration:2018}, the nominal {\it statistical} parallax errors for the Cepheids in our sample were expected to drop from $ \sim 300 \,\mu$as, typical of DR1, to $ \sim 40 \,\mu$as.  These Cepheids are all in the brightness range $6.05 < G < 11.70$ (mean magnitude; $ G $ is the natural passband of the {\Gaia} astrometric detectors).  These are fainter than the saturation limit at the shortest gating interval used (TDI gate 4, 16 lines; \citealt{Gaia-Collaboration:2016}), and thus are not expected to be sigificantly affected by saturation effects.  However, \cite[hereafter L18]{Lindegren:2018} and online material accompanying DR2\footnote{\tt https://gea.esac.esa.int/archive/documentation/GDR2/index.html} identify significant {\it systematic} uncertainties which substantially reduce the present leverage of the DR2 Cepheid parallax measurements. 

Perhaps the most significant issue with DR2 parallaxes identified in L18 is the existence of a significant parallax zeropoint error, i.e., a number which must be subtracted from all Gaia DR2 parallaxes.  In principle, large-angle astrometric measurements, such as those carried out by {\it Hipparcos} and {\Gaia}, yield an absolute parallax measurement, without the need for a correction from relative to absolute parallax. In contrast, narrow-angle parallax measurements, such as those using {\it HST} (e.g., \citealt{benedict07,riess14,Riess:2018, Casertano:2017, Brown:2018}) are only sensitive to relative parallaxes of stars within the same field, and require a correction to absolute parallax --- often based on astrophysical information.  However, as pointed out, e.g., by \cite{Michalik:2016}, instrumental uncertainties associated with monitoring the large angle between observing planes can lead to systematic errors in the determined parallaxes.  Specifically for {\Gaia}, a variation in basic angle with period equal to the spin period of the satellite is difficult to correct on the basis of self-calibration procedures; in particular, \cite{Butkevich:2017} show that the effect produced by a periodic variation of this nature is almost degenerate with a global shift of the parallaxes, resulting in a whole-sky systematic offset, i.e., a zeropoint error.  
Indeed, L18 consider the measured parallaxes for a carefully selected sample of over 500,000 quasars, whose parallaxes are expected to be extremely small, and find that they have a mean value of $ -29 \,\mu$as, with a small dependence on color and ecliptic latitude (their Fig.~7).  According to L18, ``the actual offset applicable for a given combination of magnitude, colour, and position may be different by several tens of $\mu$as''.  The quasars are primarily faint ($ G > 17 $ mag); thus, a possible magnitude dependence, suggested in their Figure 7 (left panel), is difficult to
investigate.  The distribution of corrected parallaxes $ \pi_{\rm corr} = \pi_{\rm meas} + 29 \,\mu$as is fairly consistent with a normal distribution if their nominal errors are increased by $ \sim 8\% $ (see L18, Fig. 8).  We will return to the issue of the parallax zeropoint error later in this section.

Other potential systematics identified by G18 and L18 include:

$\bullet$ Uncharacterized systematic errors dominate over the ideally available precision in the post-fit astrometric residuals (see L18, Figure 9) in DR2 by a large factor for $G < 12$ mag, with the discrepancy increasing for brighter magnitudes.  Note that a systematic deviation of parallax measurements as a function of Cepheid brightness would be somewhat degenerate with a luminosity scale determination because brightness is partially correlated with distance.

$\bullet$ A small proportion of individual parallaxes are ``corrupted''; they can generally be identified by large positive or negative values and must be discarded.
 
$\bullet$ The statistical uncertainties may be {\it underestimated} by {\it up to} $\sim$ 30 \% for stars with $G < 12$.  

We note that the spatial correlation of parallax errors on the sky for DR2 is not very significant for the Cepheids in our sample, for which only two pairs are separated by less than 10 degrees.

In light of these issues, there is likely no unique way to model the Cepheid sample while using the DR2 results to determine their luminosity scale.  Rather we take a cautious, ``common sense'' approach to illustrate what such an approach can reveal at present.  We anticipate reduction to these systematic uncertainties through independent analyses of other classes of objects and from future {\Gaia} data releases.

As a first, exploratory step we plot the DR2 parallaxes of the Cepheids in Table 1 against their uncertainties in Figure~3.  It is immediately obvious that three of the 50 (RY-Vel, RW-Cam, and SV-Per) have anomalously high formal uncertainties, and two of their parallaxes define the extrema for the set.  (Note that because of the excess noise formalism \citep{lindegren:2012}, large errors are often indicative of poor adherence to the model used, in this case a five-parameter, single-star astrometric model.)  One of the three (RW-Cam) was also an outlier in the lower precision DR1 data \citep{Casertano:2017}.  For SV-Per and RW-Cam, our {\it HST} spatial scan data demonstrate the presence of a companion within $ 0\farcs3 $ of the Cepheid; see insets in Figure~3.  Both have reported UV excess from IUE spectra consistent with B8III companions \citep{Evans:1994}.  The companions are the likely source of the anomalous astrometric solution.   All three objects are excluded from further analysis.

An additional, independent test can be carried out thanks to the existence of {\it HST} parallax measurements for 19 Cepheids, obtained using the Fine Guidance sensor \citep{benedict07} or WFC3 spatial scanning \citep{Riess:2018}.  The comparison between {\it HST} and {\Gaia} DR2 parallaxes is shown in Figure 4 vs. their mean {\Gaia} $ G $ magnitude; error bars combine the errors in the {\it HST} and DR2 parallaxes.  Two Cepheids, Y-Sgr and Delta-Cep, were excluded from this comparison because their {\Gaia} DR2 values were negative, indicating they are likely corrupted.  Delta-Cep is also a binary \citep{Anderson:2015}, and its orbit will eventually be included in the {\Gaia} solution in later releases.  For one, $\ell$-Car, its DR2 G mag was unrealistically faint, and so we plotted it at its ground-based value.  The agreement is good for Cepheids with $G>6$ mag, 7 of 8 of which fall within 1$\sigma$, but it becomes quite poor for Cepheids with $G<6$ mag (even excluding the 3 mentioned), just 1 of 11 within 1$\sigma$.  This suggests as expected that around $G \approx 6$ mag, where the {\Gaia} detectors are known to saturate, the DR2 parallaxes become much less reliable.  We conclude that even with the possible maximum 30\% enhanced {\Gaia} DR2 errors the Cepheid parallaxes at $G<6$ mag are not yet sufficiently well understood and should not be used in any quantitative analyses.  To be safe, we also exclude from further comparisons the Cepheid T-Mon: with a mean magnitude $ G \approx 6.1 $, it is the brightest in our sample (brighter by 0.3 mag in our own $F555W$ data than the next one).  Because of brightness variations in Cepheids, T-Mon is likely to have exceeded the saturation limit during some of its epochs of astrometric observations.  The comparison of the 8 Cepheids with $G>6$ mag yields a DR2 parallax offset of $-90 \pm 21 \,\mu$as but the comparison is strongly impacted by SS CMa.  Excluding SS CMa yields an offset of $-55 \pm 25 \,\mu$as. While informative and consistent with subsequent analyses, we do not make explicit use of the {\it HST} parallaxes when comparing the Gaia parallaxes to their photometric predictions to retain independence with the expectations based on \citep{Riess:2016} or \citep{Riess:2018}.

After excluding four Cepheids with too large formal uncertainties (RY-Vel, RW-Cam, and SV-Per) or too close to the saturation threshold (T-Mon), we are left with 46 Cepheids with {\it HST} photometry and reliable {\Gaia} DR2 parallaxes and uncertainties.  Following the approach of \cite{Casertano:2017}, we determine for each Cepheid the expected parallax based on its photometry and the absolute magnitude derived from the Leavitt law \citep{Leavitt:1912}, calibrated by \cite{Riess:2016} in the same photometric system.  This parallax has a typical uncertainty of only a few percent.  Figure~5 compares the measured DR2 parallax with the expected value; {\it the comparison is made in parallax space to avoid issues related to the conversion of low SNR parallaxes to magnitudes \citep{Hanson:1979} which otherwise skews their likelihood in magnitude space (see also recommendations in \citealt{Luri:2018}).}

As expected, the DR2 parallaxes are offset, on average, with respect to the predicted values.  However, a cursory examination of Figure~5, and basic statistics on the differences between predicted and measured values, suggest a zeropoint offset in the same direction but somewhat larger (in absolute value) than the value of $ -29 \,\mu$as for quasars reported by \cite{Lindegren:2018}.  Therefore, we proceed to constrain the parallax zeropoint internally from our sample, as discussed above.  Fortunately, the Cepheids in our sample have a fairly narrow range of color (a dispersion in $F555W-F814W$ of 0.28 mag) and magnitude, so we will assume zeropoint variations are small across our sample.

At the same time, we consider a possible rescaling of the photometrically-predicted distances (parallaxes) because these make direct use of the calibration of the distance ladder and attendant value of the Hubble constant, H$_0 = 73.24 \pm 1.7$ \kmsmpc, from \cite{Riess:2016}.  A degree of tension exists between this value of H$_0$ and the one determined from Planck cosmic microwave background (CMB) data in concert with the $\Lambda$-cold-darm-matter ($\Lambda$CDM) model, which yields H$_0=66.93 \pm 0.62$ \kmsmpc\ \citep{Planck-Collaboration:2016}.  To the extent the DR2 data permit an independent determination of the Cepheid luminosity calibration, they can also help distinguish between these values of H$_0$.

Therefore we seek to optimize the value of:

\bq \chi^2=\sum {( \pi_{DR2,i} - \alpha \pi_{R16,i} - zp)^2 \over \sigma_i^2}, \eq

with the two free parameters, $\alpha$ and $zp$, representing the cosmic distance scale from DR2 relative to H$_0=73.24$ {\kmsmpc} and the parallax zeropoint appropriate for the DR2 Cepheid measurements, in the direction of measured minus predicted parallaxes (consistent with the definition of L18).

To determine the individual $\sigma_i$ we add in quadrature the photometric parallax uncertainty (mean of 0.02 mag or 4 $\mu$as in parallax), the intrinsic width of the NIR Wesenheit $P-L$ (0.05 mag or a mean of 18 $\mu$as in parallax)  and the nominal parallax uncertainty as given in the DR2 release.   The mean of these $\sigma_i$ is 39 $\mu$as (median 35 $\mu$as).  

Minimizing the value of $\chi^2$ gives value of $zp =$ \facezp and $\alpha=$ \facemult, with a value of $\chi^2=$ {\facechi} for 44 degrees of freedom. Confidence regions for the two parameters are shown in Figure 6.   Although these two parameters are correlated, the range of Cepheid parallaxes, from 0.2 to 1 mas, breaks to some extent the degeneracy between $\alpha$ and $zp$ and allows for their separate determination.  

This parallax zeropoint error, somewhat different (larger in absolute value) from the value determined by L18, is significant in comparison with the formal DR2 uncertainties of our sources.  The uncertainty in the parallax zeropoint error has the potential to impact significantly any astronomical analysis based on DR2 parallaxes, especially when multiple sources are used, thus in principle reducing statistical uncertainties.  The potential dependence of the zeropoint error on source properties suggested by L18, such as color or (possibly) magnitude, is especially relevant, as it suggests that applying the nominal L18 zeropoint correction, which was determined for blue, faint objects, may not be optimal for objects with different characteristics.  Indeed, in Lindegren (2018) the offset measured from quasars appears to increase at the brighter end ($14 < G < 16$ mag) and at the redder end ($ G_{BP} \hbox{--} G_{RP} > 1$ mag), where the offset fluctuates around a higher value of $ -50 \,\mu$as, both directions that apply to Cepheids.  The online documentation also indicates that the estimated parallax zeropoint depends on the sample of sources examined \citep{Arenou:2018}, and the value determined in L18 should not be used to ``correct'' the catalogue parallax values. 

A more precise constraint on the zeropoint offset for use in other studies may be derived by fixing the value of $\alpha$ (e.g., to unity based on other geometric distance measurements to Cepheids from R16 which have a mean error of 1.4\%) which results in a constraint of \noscalezp.  That this value is more than 3$\sigma$ from the value derived from relatively bluer and fainter quasars from L18 reinforces their finding that the parallax zeropoint offset can vary with sources' position, magnitude and color, all quite different between MW Cepheids and quasars but in the right direction as suggested by the brightest and reddest quasars.

The value of $\alpha$ is quite consistent with unity, indicating that the predicted parallaxes, after accounting for the offset, are in good agrement with DR2, affirming the cosmic distance scale or the value of H$_0$ used to predict the parallaxes from \cite{Riess:2016}.  On the other hand, this value of $\alpha$ is inconsistent with the value of $\alpha=0.91$ needed to rescale the parallaxes to match the Planck CMB + $\Lambda$CDM value of H$_0$ at the \faceplanck $\sigma$ confidence level (99.6\% likelihood).  Including the 8 Cepheids with {\it HST} parallaxes from \cite{Riess:2018} to help constrain the parallax offset gives a result of $\alpha=1.035 \pm 0.029$ and $\alpha=1.010 \pm 0.029$ after excluding the one Cepheid with a large difference between {\it HST} and Gaia DR2, SS CMa.  These are 4.4 $\sigma$ and 3.4 $\sigma$ from the Planck CMB + $\Lambda$CDM value of $H_0$, respectively.

We also note that the value of $\chi^2$ appears somewhat high for the 46 Cepheids and 2 fitted parameters (44 degrees of freedom), a value which would be exceeded by chance in 3.5\% of trials.  The bottom of Figure 6 shows the residuals from the best fit versus {\Gaia} DR2 $ G $ magnitudes with a dispersion of 43 $\mu$as.  No trend is apparent nor any outliers (largest deviations is 2.3$\sigma$, expectable for 44 Cepheids and below the threshold for outlier rejection; Chauvenet's criterion would suggest a threshold of 2.6$\sigma$ for outlier rejection for a sample with 46 objects).  

If we consider the high $\chi^2$ to be an indication of additional variance in the data, a promising source is suggested by L18 and other material accompanying DR2, which states that for bright targets ($G<12$ mag) formal errors may be underestimated by up to 30\%.  Rescaling the DR2 parallax errors by ${\chi^2_{\rm dof}}^{1/2} \approx 1.19$ raises the mean error to 46 $\mu$as.  We refit the model and find $zp =$ {\expandzp} and $\alpha=$ {\expandmult} with a value of $\chi^2=$ {\expandchi}; the inconsistency with Planck CMB + $\Lambda$CDM is \expandplanck$\sigma$.  For the expanded errors there is now no Cepheid with a deviation $> 2\sigma$.  Because we would expect between 2 and 3 such Cepheids, one might argue that the expanded errors are now {\it too} large.  However, we think this identifies the range of reasonability for fitting these data. 

Unfortunately, the {\it cost} of needing to measure the appropriate zeropoint offset from the Cepheid sample is (painfully) large.  The marginalized uncertainty in $\alpha$ is 0.033, providing a 3.3\% independent calibration of the cosmic distance scale, 2.5 times what would otherwise result from the formal parallax uncertainties and full knowledge of the parallax zeropoint, {\it better} than it has ever been determined in the local Universe.  As an illustration, in Figure 8 we use the constraint of -53 $\pm 2.6$ $\mu$as on the parallax zeropoint offset calculated from 3475 Red Giants with Kepler-based asteroseismic estimates of radii and parallaxes from \cite{Zinn:2018}.  The mean color of this Red Giant sample well matches the Cepheids (greater optical extinction of the Cepheids compensates their bluer color). The Red Giant mean magnitudes are a few magnitudes fainter than the Cepheid mean but much closer than the Quasar sample used by \cite{Lindegren:2018}.  It is therefore not surprising that the constraint from the Red Giants is quite consistent with the Cepheids.   We have chosen not to formally include it in the determination of $H_0$ due to its model-dependence.  However, this external constraint, which is five times more precise than the internal constraint, demonstrates the value of independent knowledge of the DR2 offset term.  Making use of it reduces the uncertainty in the distance scale for the {\it HST} Cepheid sample to $\sim$ 1.3\%.  
  
We are optimistic that future {\Gaia} data releases will resolve the uncertainty of the parallax zeropoint offset while producing parallax measurements near the expectations for the end of the mission.  With these and the {\it HST} photometry presented here we would expect to reach the full potential precision of $\sim$ 0.5\% from this Cepheid sample.  

\subsection{ Ground-Based Sample, Caveats}  

To improve the constraint on $zp$ and $\alpha$ we might consider using a larger sample of MW Cepheids, though the augmentation of the sample would need to rely exclusively on ground-based photometry.  There are compilations of ground-based Cepheid photometry which could augment the {\it HST} sample by an additional $\sim $ 150-250 Cepheids, for example from \cite{vanleeuwen07}.  

However, because the {\it HST} sample of 50 presented here were selected to have $P>$ 8 days, $A_H < 0.4 $ mag, $V > 6$ mag, $D < 6$ kpc and is largely ($>$ 70\%) complete (with selection made randomly by the scheduling of {\it HST}), an expanded sample would be dominated by Cepheids which necessarily violate these criteria.  Most would have $A_H > 0.4$ or $P<$ 8 days with resulting negative consequences.  

On average, such shorter period Cepheids are bluer in their mean $B-V$ by $\sim$ 0.2 mag which might alter their parallax zeropoints relative to the redder {\it HST} sample.  In addition, these Cepheids are a couple of magnitudes fainter, so that their astrometric observations are more easily contaminated by a companion. 

Further, Cepheids with $P < 8$ days would be shorter than the period range they would be used to calibrate, i.e., those which are visible in distant SN Ia hosts, putting too great a reliance on the linearity of the {\PL} relation.  Adding Cepheids with $A_H> 0.4$ would lead to larger magnitude errors due to variations in the reddening law.  

Additional loss in precision is expected from the use of ground-based photometry in lieu of {\it HST} photometry for the expanded sample.  Ground-based photometry covering two hemispheres is by necessity quite inhomogeneous, especially in the NIR where limited standards are available.  The few truly wide-angle surveys to date in the NIR lack time sampling needed to determine the mean of the light curves.  Moreover, in the NIR, the correspondance between the ground-based $H$-band filter and the WFC3 $F160W$ filter is particularly poor as indicated by the large color term of $\sim 0.2$ mag per mag of $J-H$ color which has been measured between the two \citep{Riess:2016, riess11c}.  We find systematic errors are readily apparent between different sources of ground-based NIR Cepheid photometry.  A comparison of the 79 Cepheids in common between the two most recent compilations, \cite{Monson+2011} and \cite{vanleeuwen07} shows a gradient of 0.015 mag per mag at $>3\sigma$ confidence in $H$ and a mean difference of 0.08 mag in $J$.  
  
We also note that the comparison of the {\it HST} system photometry for 50 Cepheids presented here with ground equivalent (see Appendix for source) indicated {\it a systematic difference of -0.051 mag} (Ground-{\it HST}) in the Wesenheit magnitudes used to measure distances.  The size of this systematic error would likely depend on the specific ground-based system used, or their mixture for heterogeneous collections.

Lastly, without the use of the high resolution {\it HST} data one would lose the means to test for contamination of parallaxes by nearby companions as illustrated in two of three cases in \S~2.  
 
To better compare the size of the uncertainties associated with {\it HST} Cepheid sample and an augmentation to it from ground-based data we produced a Cepheid sample comprised of all the Cepheids with photometry from a single source; NIR mean magnitudes from \cite{Monson+2011} for Northern Cepheids and $V$ and $I$-band mean magnitudes from \cite{Berdnikov+2000}, excluding objects in the {\it HST} sample, leaving 86 additional Cepheids.   For these we included the -0.051 mag offset identified between the ground and {\it HST} measurements of $m^W_H$ and the reduced CRNL of 0.036 mag (reduced to the 4.5 dex that applies from extragalactic Cepheids to {\it HST} system standards; see notes in Table 1) that would apply between the ground and extragalactic Cepheids.  A basic comparison to the DR2 parallaxes is shown in Figure 7.  The augmented sample, though in rough agreement with the {\it HST} sample, has far greater errors with a dispersion of differences (after removing the parallax zeropoint offset) of 99 $\mu$as, 2.3 times that of the {\it HST} sample.  Even removing the most deviant points leaves a high dispersion of 60-070 $\mu$as.  This level of uncertainty is far greater than we can model by increasing the {\Gaia} DR2 parallax errors, even by the maximum suggested value of 30\% as it would require $\sim$ 70\%.  It is hard to realistically characterize the source of this additional variance and whether it may belie other important dependencies and we therefore decided to not make further use of an augmented sample.

\section{Discussion}

\subsection{The Zeropoint Error in DR2 Parallaxes}

We have presented an analysis of the {\Gaia} DR2 parallax values and their uncertainties for a carefully selected sample of 50 Cepheids with precise, consistent photometry obtained with {\it HST} using spatial scanning observations.  The photometry for most of these Cepheids is published here for the first time (Table 1).  The high accuracy of the Leavitt law calibration for these Cepheids, obtained by R16 on the basis of several independent anchors, allows us to predict their parallaxes with uncertainties much smaller than those of the DR2 parallaxes.  We also consider a larger sample of Cepheids covering a broader range of magnitudes and properties but with only ground-based photometry, and a small sample of Cepheids for which {\it HST} parallaxes have been published.

Our first conclusion is that the {\Gaia} DR2 parallaxes for our Cepheids, in the {\Gaia} magnitude range $ 6 < G < 12 $ mag, are generally in good agreement with their predicted values.  We confirm the existence of a zeropoint parallax error indicated in the {\Gaia} release material; however, we find a somewhat larger (more negative) value for the zeropoint, \facezp, where the uncertainty includes marginalizing over the possible recalibration of the Leavitt law on the basis of DR2 parallaxes alone.  Using the R16 period-luminosity calibration without rescaling, the value of the parallax zeropoint inferred from this sample of Cepheids is \noscalezp.  The difference between our estimated zeropoint and the value obtained by L18 from quasars suggests that the zeropoint does depend on magnitude, color, or position on the sky (all are different for Cepheids), as suggested by L18; online DR2 documentation similarly states that the zeropoint depends on the sample used.  We emphasize the need to include zeropoint uncertainties in any analysis based on DR2 parallaxes; an independent determination of the parallax zeropoint should be carried out for any data for which this is possible --- for example, via asteroseismology \citep{De-Ridder:2016} and eclipsing binaries \citep{Stassun:2016}.  We also suggest a possible increase of the formal DR2 errors for stars in this range by about 19\%, with modest (96.5\%) significance.

Comparison of DR2 with {\it HST} parallaxes suggest that parallaxes for bright stars ($ G < 6  $) may be unreliable, consistent with the large residuals L18 find for bright stars.  This conclusion is reinforced by the analysis of the larger sample of Cepehids with ground-based photometry.  We also find that at the level of precision of DR2 parallaxes, existing ground-based photometry is of insufficient quality to take full advantage of the parallax information; photometric errors are likely underestimated, possibly because of systematic offsets between systems and between standards in different parts of the sky.  This will be even more true with future releases, when {\Gaia} precision is expected to improve significantly, and zeropoint issues will likely be addressed.  We would recommend that only Cepheids with accurate, high-quality photometry, free of systematic effects, should be used in the calibration of the Leavitt law with the precision enabled by {\Gaia} DR2 and beyond.

\subsection{Implications for Determination of the Hubble Constant}

  The results presented here may be evaluated as (another) independent test of the scale of the local determination of H$_0$ from \cite{Riess:2016} or as an augmentation to that measurement.  As an independent test, the results from constraining $\alpha$ reaffirm the present ``tension'' between the local determination of H$_0$ and that based on Planck CMB data in concert with $\Lambda$CDM \citep{Planck-Collaboration:2016}.  This test is similar in outcome to the one from \cite{Riess:2018} which employed measurements of 8 parallaxes of long-period Cepheids using spatial scanning on {\it HST} and reaching a mean precision of 45 $\mu$as, similar to the {\Gaia} DR2 formal precision.  The key differences are that the present study uses a factor of 5 times as many Cepheids but that statistical advantage is largely returned by the need to determine the offset in the {\Gaia} DR2 Cepheid parallaxes.  
  
  By including the new MW parallaxes from {\it HST} and {\Gaia} to the rest of the data from \cite{Riess:2016} the value of H$_0$ changes slightly to 73.52 $\pm$ 1.62 (including systematics discussed in R16) and increases the tension to 3.8$\sigma$.  While we have chosen not to formally use the \cite{Zinn:2018} external constraint on the parallax offset based on Red Giants due to its model-dependence, we note including it would result in $H_0=73.83 \pm 1.48$ and would raise tension to 4.3 $\sigma$, thus illustrating the leverage that such knowledge of the offset provides.  
  
  Undoubtedly the greater benefit derived from these two new sets of of parallaxes is as independent tests of luminosity calibration derived from the masers in NGC 4258 \citep{humphreys13,Riess:2016}, the detached eclipsing binaries in the LMC \citep{Pietrzynski:2013}, and shorter period, nearer MW parallaxes \citep{benedict07, vanleeuwen07}.  It is very difficult to imagine an unknown significant systematic error which would affect all 5 sources of Cepheid luminosity calibration to a comparable level.  
  
  With improved parallaxes from {\Gaia} in the future and better knowledge of their zeropoint and with observations of Cepheids in new hosts of Type Ia supernovae (now underway), a target precision for H$_0$ of $\sim 1$\% is not out of reach and would be an invaluable aid for resolving the source of the present tension.

\bigskip

\bigskip

\acknowledgements

We are grateful to the entire {\Gaia} collaboration for providing data and assistance which made this project possible.  We congratulate them on their tremendous achievement to date.

Support for this work was provided by the National Aeronautics and Space Administration (NASA) through programs GO-12879, 13334, 13335, 13344, 13571, 13678, 13686, 13928, 13929, 14062, 14394, 14648, 14868 from the Space Telescope Science Institute (STScI), which is operated by AURA, Inc., under NASA contract NAS 5-26555. B.B. and M.G.L. are indebted to the Italian Space Agency (ASI) for their continuing support through contract
2014-025-R.1.2015 to INAF, and to R. Morbidelli (INAF) and R. Messineo (ALTEC S.p.A, Torino) of the Italian Data Processing Center (DPCT) for their help with the {\Gaia} data.  A.V.F.’s group at UC Berkeley is also grateful for financial assistance from the TABASGO Foundation, the Christopher R. Redlich Fund, and the Miller Institute for Basic Research in Science (UC Berkeley). S.C. and A.G.R. gratefully acknowledge support by the Munich Institute for Astro- and Particle Physics (MIAPP) of the DFG cluster of excellence ``Origin and Structure of the Universe.'' 

This research is based primmarily on observations with the NASA/ESA {\it Hubble Space Telescope}, obtained at STScI, which is operated by AURA, Inc., under NASA contract NAS 5-26555. This publication makes use of data products from the {\it Wide-field Infrared Survey Explorer (WISE)}, which is a joint project of the University of California (Los Angeles) and the Jet Propulsion Laboratory/California Institute of Technology, funded by NASA. It has also made use of the SIMBAD database, operated at CDS, Strasbourg, France. 
Research at Lick Observatory is partially supported by a generous gift from Google.

The {\it HST} data used in this paper are available at the MAST archive
 http://dx.doi.org/10.17909/T9G40B

  \appendix
  
  {\bf Sources for Ground-based Phase Corrections}

    Compared to the \citet{Riess:2018} analysis, we make use of additional ground measurements from the ASAS-SN \citep{2014ApJ...788...48S} web interface \citep{2017PASP..129j4502K}, \citet{Berdnikov+2000}, \citet{2007PZP.....7...32B}, \citet{2015yCat..90410027B}, and \citet{vanleeuwen07}.

\begin{deluxetable}{lccccc}
\setlength\tabcolsep{0.4cm}
\def\arraystretch{0.9}
\tablenum{2}
\tablecaption{Ground Data Sources\label{tbl_obj_src}}
\tablehead{
\colhead{Identifier} & \multicolumn5c{References$^a$} \\
\cline{2-6}
& \colhead{Phase determination} & \colhead{$V$} & \colhead{$I$} & \colhead{$J$} & \colhead{$H$}
}
\startdata
AA Gem & 1,2,5,10,11,13-15,18,20 & 5,10 & 11 & 19 & 19 \\
AD Pup & 2,3,5,7-11,18,31 & 5,10,31 & 2,3,11 & NA & NA \\
AQ Car & 2,7,10,11,35 & 10 & 2,11,35 & NA & NA \\
AQ Pup & 1-4,6,7,9,10,13,18,21,29-31 & 1-3,9,13,18,21 & 2,3,21 & 4 & 4 \\
BK Aur & 2,25,27 & 2,25,27 & NA & NA & NA \\
BN Pup & 2,4-7,9-11,18,21,29,31 & 5,10,31 & 2,11,21 & 4 & 4 \\
CD Cyg & 1,2,5,7,13,15,16,18-20,32 & 1,2,7,13,15,16,18,20 & 13 & 19,32 & 19 \\
CP Cep & 2,5,7,14,16,18 & 5 & NA & 19 & 19 \\
CR Car & 2,5,8,11,33 & 5 & 2,11 & NA & NA \\
CY Aur & 1,2,5,7,18,38 & 5 & NA & 19 & 19 \\
DD Cas & 2,5,7,13,15,24,25 & 5 & 13 & 19 & 19 \\
DL Cas & 1,2,7,13-19,23-26 & 1,2,7,13-18,23-26 & 13 & 19 & 19 \\
DR Vel & 7,9-11,21 & 7,9-11,21 & 11,21 & NA & NA \\
GQ Ori & 2,11,13,25,41 & 41 & 41 & 42 & 42 \\
HW Car & 2,7,10,11 & 2,7,10,11 & 2,11 & NA & NA \\
KK Cen & 2,5,9-11,33 & 5 & 2,11 & NA & NA \\
KN Cen & 1-12 & 5,10 & 2,3,11 & 4 & 4 \\
RW Cam & 1,2,5,7,13,14,16,18,20,36 & 1,7,13,14,16,18,20,36 & 13 & 19 & 19 \\
RW Cas & 1,2,5,7,13,14,16,18,20 & 1,2,5,7,13,14,16,18,20 & 13 & 19 & 19 \\
RY Cas & 1,2,5,7,14,18 & 5 & NA & 19 & 19 \\
RY Sco & 1,2,4,6,9,10,13,21,32 & 1,2,9,13,21 & 2,21 & 4 & 4 \\
RY Vel & 1-4,6-10,32 & 10 & 2,3 & 4 & 4 \\
S Nor & 2,4,6,32 & 2 & 2 & 4 & 4 \\
S Vul & 19,31,40 & 40 & NA & 19 & 19 \\
SS CMa & 2,5,7,9-11,18,21,31 & 5,10,31 & 2,11,21 & NA & NA \\
SV Per & 1,2,7,13,14,18,20,36 & 1,2,7,13,14,18,20,36 & 13 & 19 & 19 \\
SV Vel & 2,3,7,9,10 & 10 & 2,3 & NA & NA \\
SV Vul & 2,5,15,16,28,34 & 5 & 2 & 28 & 4,19 \\
SY Nor & 1,2,5,7-11,33 & 5,10 & 2,11 & NA & NA \\
SZ Cyg & 1,2,7,13,14,18,20 & 1,2,7,13,14,18,20 & 13 & 19 & 19 \\
T Mon & 1-6,13-22 & 1-3,13-18,20-22 & 2,3,21 & 4 & 4,6,19 \\
U Car & 2-6,21,32,33 & 2,3,5,21,33 & 2,3,21 & 4 & 4,6,32 \\
UU Mus & 4-7,9-11,21 & 5,10 & 11,21 & 4 & 4 \\
V339 Cen & 2,10,11,21 & 2,10,11,21 & 2,11,21 & NA & NA \\
V340 Ara & 1,2,5,7,9-11 & 5,10 & 2,11 & NA & NA \\
VW Cen & 1,2,4-8,10,11,21 & 5,10 & 2,11,21 & 4 & 4 \\
VX Per & 1,2,7,13,15,20,26,36,37 & 1,2,7,13,15,20,26,36,37 & 13 & 19 & 19 \\
VY Car & 1,2,4-6,9,21,32,39 & 1,2,5,9,21,39 & 2,21,39 & 4 & 4 \\
VZ Pup & 1-11,18,21,29-31 & 1-3,7,9-11,18,21,31 & 2,3,11,21 & 4 & 4 \\
WX Pup & 2,7,8,10,11,13,31 & 10,31 & 2,11 & NA & NA \\
WZ Sgr & 1,2,4,6,7,9,10,13,14,18,19,21 & 10 & 2,21 & 4,19 & 4,6,19 \\
X Cyg & 1,2,5,13-18,20,23,27,28,32,34 & 1,2,13-18,20,23,27,28,34 & 2 & 28,32 & NA \\
X Pup & 1-4,6-10,13,18,30 & 10 & 2,3 & 4 & 4 \\
XX Car & 2,7,9-11,35 & 10 & 2,11,35 & NA & NA \\
XY Car & 1,7,9-11,35 & 10 & 11,35 & NA & NA \\
XZ Car & 1,2,7,9,10,35 & 10 & 2,35 & NA & NA \\
YZ Car & 1,2,7,9-11,21 & 10 & 2,11,21 & NA & NA \\
YZ Sgr & 2,3,6,10,13,15-17,19,23,24,27,32,33 & 10 & 2,3 & 19,32 & 6,19 \\
Z Lac & 1,2,7,13-20,23,27,28 & 1,2,7,13-18,20,23,27,28 & 13,28 & 19,28 & NA \\
Z Sct & 1,2,5,7-10,13,14,18,31 & 5,10,31 & 2 & NA & NA \\
\enddata

\tablecomments{
$^a$ The labels are described in Table~\ref{tbl_ref_src}. NA indicates no ground data avaliable.\\
}
\setlength\tabcolsep{6pt}
\def\arraystretch{1}
\end{deluxetable}

\begin{deluxetable}{lcc}
\setlength\tabcolsep{0.4cm}
\def\arraystretch{0.9}
\tablenum{3}
\tablecaption{References for the Labels in Table~\ref{tbl_obj_src}\label{tbl_ref_src}}
\tablehead{
\colhead{Reference ID} & \colhead{Reference} & \colhead{Comments}
}
\startdata
1 & \citet{Harris1980} & McMaster\\
2 & \citet{Berdnikov+2000} & \\
3 & \citet{Berdnikov+1995} & McMaster\\
4 & \citet{Laney+1992} & McMaster\\
5 & \citet{2017PASP..129j4502K} & ASAS-SN\\
6 & This work & CTIO observations\\
7 & \citet{Alfonso-Garzon+2012} & I-OMC\\
8 & \citet{Pel1976} & McMaster\\
9 & \citet{Madore1975} & McMaster\\
10 & \citet{Pojmanski1997} & ASAS\\
11 & \citet{2015yCat..90410027B} & \\
12 & \citet{Walraven+1964} & McMaster\\
13 & \citet{Moffett+1984} & McMaster\\
14 & \citet{1992AandAT....2...43B} & McMaster\\
15 & \citet{Berdnikov1992} & McMaster\\
16 & \citet{1993AstL...19...84B} & McMaster\\
17 & \citet{1995AstL...21..308B} & McMaster\\
18 & \citet{1986PZ.....22..369B} & McMaster\\
19 & \citet{Monson+2011} & \\
20 & \citet{Szabados1981} & McMaster\\
21 & \citet{Coulson+1985} & McMaster\\
22 & \citet{1992AandAT....2...31B} & McMaster\\
23 & \citet{1992AandAT....2..107B} & McMaster\\
24 & \citet{1992AandAT....2....1B} & McMaster\\
25 & \citet{Szabados1980} & McMaster\\
26 & \citet{Berdnikov1987} & McMaster\\
27 & \citet{1992AandAT....2..157B} & McMaster\\
28 & \citet{Barnes+1997} & McMaster\\
29 & \citet{Schechter+1992} & McMaster\\
30 & \citet{Welch1986} & McMaster\\
31 & Dorota Szczygiel, private communication & ASAS (unreleased)\\
32 & \citet{Welch+1984} & McMaster\\
33 & \citet{Berdnikov+1995} & McMaster\\
34 & \citet{Kiss1998} & McMaster\\
35 & \citet{Coulson+1985b} & McMaster\\
36 & \citet{Szabados1991} & McMaster\\
37 & \citet{Berdnikov+1993} & McMaster\\
38 & \citet{Schmidt+1995} & McMaster\\
39 & \citet{2007PZP.....7...32B} & \\
40 & This work & RCT observations\\
41 & This work & Lick-Kait observations\\
42 & \citet{vanleeuwen07} & Mean magnitude\\
\enddata
\setlength\tabcolsep{6pt}
\def\arraystretch{1}
\end{deluxetable}

\vfill
\eject

\clearpage
\bibliographystyle{aasjournal} %
\bibliography{bibdesk}

\begin{thebibliography}{}
\expandafter\ifx\csname natexlab\endcsname\relax\def\natexlab#1{#1}\fi
\providecommand{\url}[1]{\href{#1}{#1}}

\bibitem[{{Alfonso-Garz{\'o}n} {et~al.}(2012){Alfonso-Garz{\'o}n}, {Domingo},
  {Mas-Hesse}, \& {Gim{\'e}nez}}]{Alfonso-Garzon+2012}
{Alfonso-Garz{\'o}n}, J., {Domingo}, A., {Mas-Hesse}, J.~M., \& {Gim{\'e}nez},
  A. 2012, \aap, 548, A79

\bibitem[{{Anderson} \& {Riess}(2017)}]{Anderson:2017}
{Anderson}, R.~I., \& {Riess}, A.~G. 2017, ArXiv e-prints, arXiv:1712.01065

\bibitem[{{Anderson} {et~al.}(2015){Anderson}, {Sahlmann}, {Holl}, {Eyer},
  {Palaversa}, {Mowlavi}, {S{\"u}veges}, \& {Roelens}}]{Anderson:2015}
{Anderson}, R.~I., {Sahlmann}, J., {Holl}, B., {et~al.} 2015, \apj, 804, 144

\bibitem[{{Arenou} {et~al.}(2018){Arenou}, {Luri}, {Babusiaux}, {Fabricius},
  {Helmi}, {Muraveva}, {Robin}, {Spoto}, {Vallenari}, {Antoja},
  {Cantat-Gaudin}, {Jordi}, {Leclerc}, {Reyl{\'e}}, {Romero-G{\'o}mez}, {Shih},
  {Soria}, {Barache}, {Bossini}, {Bragaglia}, {Breddels}, {Fabrizio},
  {Lambert}, {Marrese}, {Massari}, {Moitinho}, {Robichon}, {Ruiz-Dern},
  {Sordo}, {Veljanoski}, {Di Matteo}, {Eyer}, {Jasniewicz}, {Pancino},
  {Soubiran}, {Spagna}, {Tanga}, {Turon}, \& {Zurbach}}]{Arenou:2018}
{Arenou}, F., {Luri}, X., {Babusiaux}, C., {et~al.} 2018, ArXiv e-prints,
  arXiv:1804.09375

\bibitem[{{Barnes} {et~al.}(1997){Barnes}, {Fernley}, {Frueh}, {Navas},
  {Moffett}, \& {Skillen}}]{Barnes+1997}
{Barnes}, III, T.~G., {Fernley}, J.~A., {Frueh}, M.~L., {et~al.} 1997, \pasp,
  109, 645

\bibitem[{{Benedict} {et~al.}(2007){Benedict}, {McArthur}, {Feast}, {Barnes},
  {Harrison}, {Patterson}, {Menzies}, {Bean}, \& {Freedman}}]{benedict07}
{Benedict}, G.~F., {McArthur}, B.~E., {Feast}, M.~W., {et~al.} 2007, \aj, 133,
  1810

\bibitem[{{Berdnikov}(1986)}]{1986PZ.....22..369B}
{Berdnikov}, L.~N. 1986, Peremennye Zvezdy, 22, 369

\bibitem[{{Berdnikov}(1987)}]{Berdnikov1987}
---. 1987, Peremennye Zvezdy, 22, 530

\bibitem[{{Berdnikov}(1992{\natexlab{a}})}]{1992AandAT....2...43B}
---. 1992{\natexlab{a}}, Astronomical and Astrophysical Transactions, 2, 43

\bibitem[{{Berdnikov}(1992{\natexlab{b}})}]{Berdnikov1992}
---. 1992{\natexlab{b}}, Soviet Astronomy Letters, 18, 130

\bibitem[{{Berdnikov}(1992{\natexlab{c}})}]{1992AandAT....2...31B}
---. 1992{\natexlab{c}}, Astronomical and Astrophysical Transactions, 2, 31

\bibitem[{{Berdnikov}(1992{\natexlab{d}})}]{1992AandAT....2..107B}
---. 1992{\natexlab{d}}, Astronomical and Astrophysical Transactions, 2, 107

\bibitem[{{Berdnikov}(1992{\natexlab{e}})}]{1992AandAT....2....1B}
---. 1992{\natexlab{e}}, Astronomical and Astrophysical Transactions, 2, 1

\bibitem[{{Berdnikov}(1992{\natexlab{f}})}]{1992AandAT....2..157B}
---. 1992{\natexlab{f}}, Astronomical and Astrophysical Transactions, 2, 157

\bibitem[{{Berdnikov}(1993)}]{1993AstL...19...84B}
---. 1993, Astronomy Letters, 19, 84

\bibitem[{{Berdnikov} {et~al.}(2000){Berdnikov}, {Dambis}, \&
  {Vozyakova}}]{Berdnikov+2000}
{Berdnikov}, L.~N., {Dambis}, A.~K., \& {Vozyakova}, O.~V. 2000, \aaps, 143,
  211

\bibitem[{{Berdnikov} {et~al.}(2015){Berdnikov}, {Kniazev}, {Sefako}, {Dambis},
  {Kravtsov}, \& {Zhuiko}}]{2015yCat..90410027B}
{Berdnikov}, L.~N., {Kniazev}, A.~Y., {Sefako}, R., {et~al.} 2015, VizieR
  Online Data Catalog, 904

\bibitem[{{Berdnikov} {et~al.}(2007){Berdnikov}, {Kravtsov}, {Pastukhova}, \&
  {Turner}}]{2007PZP.....7...32B}
{Berdnikov}, L.~N., {Kravtsov}, V.~V., {Pastukhova}, E.~N., \& {Turner}, D.~G.
  2007, Peremennye Zvezdy Prilozhenie, 7

\bibitem[{{Berdnikov} \& {Turner}(1995)}]{Berdnikov+1995}
{Berdnikov}, L.~N., \& {Turner}, D.~G. 1995, Astronomy Letters, 21, 717

\bibitem[{{Berdnikov} \& {Vozyakova}(1995)}]{1995AstL...21..308B}
{Berdnikov}, L.~N., \& {Vozyakova}, O.~V. 1995, Astronomy Letters, 21, 308

\bibitem[{{Berdnikov} \& {Yakubov}(1993)}]{Berdnikov+1993}
{Berdnikov}, L.~N., \& {Yakubov}, S.~D. 1993, Peremennye Zvezdy, 23, 47

\bibitem[{{Brown} {et~al.}(2018){Brown}, {Casertano}, {Strader}, {Riess},
  {VandenBerg}, {Soderblom}, {Kalirai}, \& {Salinas}}]{Brown:2018}
{Brown}, T.~M., {Casertano}, S., {Strader}, J., {et~al.} 2018, \apjl, 856, L6

\bibitem[{{Butkevich} {et~al.}(2017){Butkevich}, {Klioner}, {Lindegren},
  {Hobbs}, \& {van Leeuwen}}]{Butkevich:2017}
{Butkevich}, A.~G., {Klioner}, S.~A., {Lindegren}, L., {Hobbs}, D., \& {van
  Leeuwen}, F. 2017, \aap, 603, A45

\bibitem[{{Casertano} {et~al.}(2017){Casertano}, {Riess}, {Bucciarelli}, \&
  {Lattanzi}}]{Casertano:2017}
{Casertano}, S., {Riess}, A.~G., {Bucciarelli}, B., \& {Lattanzi}, M.~G. 2017,
  \aap, 599, A67

\bibitem[{{Casertano} {et~al.}(2016){Casertano}, {Riess}, {Anderson},
  {Anderson}, {Bowers}, {Clubb}, {Cukierman}, {Filippenko}, {Graham},
  {MacKenty}, {Melis}, {Tucker}, \& {Upadhya}}]{Casertano:2016}
{Casertano}, S., {Riess}, A.~G., {Anderson}, J., {et~al.} 2016, \apj, 825, 11

\bibitem[{{Coulson} \& {Caldwell}(1985)}]{Coulson+1985}
{Coulson}, I.~M., \& {Caldwell}, J.~A.~R. 1985, South African Astronomical
  Observatory Circular, 9

\bibitem[{{Coulson} {et~al.}(1985){Coulson}, {Caldwell}, \&
  {Gieren}}]{Coulson+1985b}
{Coulson}, I.~M., {Caldwell}, J.~A.~R., \& {Gieren}, W.~P. 1985, \apjs, 57, 595

\bibitem[{{De Ridder} {et~al.}(2016){De Ridder}, {Molenberghs}, {Eyer}, \&
  {Aerts}}]{De-Ridder:2016}
{De Ridder}, J., {Molenberghs}, G., {Eyer}, L., \& {Aerts}, C. 2016, \aap, 595,
  L3

\bibitem[{{Evans}(1994)}]{Evans:1994}
{Evans}, N.~R. 1994, \apj, 436, 273

\bibitem[{{Gaia Collaboration} {et~al.}(2018){Gaia Collaboration}, {Brown},
  {Vallenari}, {Prusti}, {de Bruijne}, {Babusiaux}, \&
  {Bailer-Jones}}]{Gaia-Collaboration:2018}
{Gaia Collaboration}, {Brown}, A.~G.~A., {Vallenari}, A., {et~al.} 2018, ArXiv
  e-prints, arXiv:1804.09365

\bibitem[{{Gaia Collaboration} {et~al.}(2016{\natexlab{a}}){Gaia
  Collaboration}, {Prusti}, {de Bruijne}, {Brown}, {Vallenari}, {Babusiaux},
  {Bailer-Jones}, {Bastian}, {Biermann}, {Evans}, \&
  et~al.}]{Gaia-Collaboration:2016}
{Gaia Collaboration}, {Prusti}, T., {de Bruijne}, J.~H.~J., {et~al.}
  2016{\natexlab{a}}, \aap, 595, A1

\bibitem[{{Gaia Collaboration} {et~al.}(2016{\natexlab{b}}){Gaia
  Collaboration}, {Brown}, {Vallenari}, {Prusti}, {de Bruijne}, {Mignard},
  {Drimmel}, {Babusiaux}, {Bailer-Jones}, {Bastian}, \&
  et~al.}]{Gaia-Collaboration:2016a}
{Gaia Collaboration}, {Brown}, A.~G.~A., {Vallenari}, A., {et~al.}
  2016{\natexlab{b}}, \aap, 595, A2

\bibitem[{{Hanson}(1979)}]{Hanson:1979}
{Hanson}, R.~B. 1979, \mnras, 186, 875

\bibitem[{{Harris}(1980)}]{Harris1980}
{Harris}, H.~C. 1980, PhD thesis, Washington Univ., Seattle.

\bibitem[{{Hoffmann} {et~al.}(2016){Hoffmann}, {Macri}, {Riess}, {Yuan},
  {Casertano}, {Foley}, {Filippenko}, {Tucker}, {Chornock}, {Silverman},
  {Welch}, {Goobar}, \& {Amanullah}}]{Hoffmann:2016}
{Hoffmann}, S.~L., {Macri}, L.~M., {Riess}, A.~G., {et~al.} 2016, \apj, 830, 10

\bibitem[{{Humphreys} {et~al.}(2013){Humphreys}, {Reid}, {Moran}, {Greenhill},
  \& {Argon}}]{humphreys13}
{Humphreys}, E.~M.~L., {Reid}, M.~J., {Moran}, J.~M., {Greenhill}, L.~J., \&
  {Argon}, A.~L. 2013, \apj, 775, 13

\bibitem[{{Kiss}(1998)}]{Kiss1998}
{Kiss}, L.~L. 1998, \mnras, 297, 825

\bibitem[{{Kochanek} {et~al.}(2017){Kochanek}, {Shappee}, {Stanek}, {Holoien},
  {Thompson}, {Prieto}, {Dong}, {Shields}, {Will}, {Britt}, {Perzanowski}, \&
  {Pojma{\'n}ski}}]{2017PASP..129j4502K}
{Kochanek}, C.~S., {Shappee}, B.~J., {Stanek}, K.~Z., {et~al.} 2017, \pasp,
  129, 104502

\bibitem[{{Laney} \& {Stobie}(1992)}]{Laney+1992}
{Laney}, C.~D., \& {Stobie}, R.~S. 1992, \aaps, 93, 93

\bibitem[{{Leavitt} \& {Pickering}(1912)}]{Leavitt:1912}
{Leavitt}, H.~S., \& {Pickering}, E.~C. 1912, Harvard College Observatory
  Circular, 173, 1

\bibitem[{{Lindegren} {et~al.}(2012){Lindegren}, {Lammers}, {Hobbs},
  {O'Mullane}, {Bastian}, \& {Hern{\'a}ndez}}]{lindegren:2012}
{Lindegren}, L., {Lammers}, U., {Hobbs}, D., {et~al.} 2012, \aap, 538, A78

\bibitem[{{Lindegren} {et~al.}(2018){Lindegren}, {Hernandez}, {Bombrun},
  {Klioner}, {Bastian}, {Ramos-Lerate}, {de Torres}, {Steidelmuller},
  {Stephenson}, {Hobbs}, {Lammers}, {Biermann}, {Geyer}, {Hilger}, {Michalik},
  {Stampa}, {McMillan}, {Castaneda}, {Clotet}, {Comoretto}, {Davidson},
  {Fabricius}, {Gracia}, {Hambly}, {Hutton}, {Mora}, {Portell}, {van Leeuwen},
  {Abbas}, {Abreu}, {Altmann}, {Andrei}, {Anglada}, {Balaguer-Nunez},
  {Barache}, {Becciani}, {Bertone}, {Bianchi}, {Bouquillon}, {Bourda},
  {Brusemeister}, {Bucciarelli}, {Busonero}, {Buzzi}, {Cancelliere},
  {Carlucci}, {Charlot}, {Cheek}, {Crosta}, {Crowley}, {de Bruijne}, {de
  Felice}, {Drimmel}, {Esquej}, {Fienga}, {Fraile}, {Gai}, {Garralda},
  {Gonzalez-Vidal}, {Guerra}, {Hauser}, {Hofmann}, {Holl}, {Jordan},
  {Lattanzi}, {Lenhardt}, {Liao}, {Licata}, {Lister}, {Loffler}, {Marchant},
  {Martin-Fleitas}, {Messineo}, {Mignard}, {Morbidelli}, {Poggio}, {Riva},
  {Rowell}, {Salguero}, {Sarasso}, {Sciacca}, {Siddiqui}, {Smart}, {Spagna},
  {Steele}, {Taris}, {Torra}, {van Elteren}, {van Reeven}, \&
  {Vecchiato}}]{Lindegren:2018}
{Lindegren}, L., {Hernandez}, J., {Bombrun}, A., {et~al.} 2018, ArXiv e-prints,
  arXiv:1804.09366

\bibitem[{{Luri} {et~al.}(2018){Luri}, {Brown}, {Sarro}, {Arenou},
  {Bailer-Jones}, {Castro-Ginard}, {de Bruijne}, {Prusti}, {Babusiaux}, \&
  {Delgado}}]{Luri:2018}
{Luri}, X., {Brown}, A.~G.~A., {Sarro}, L.~M., {et~al.} 2018, ArXiv e-prints,
  arXiv:1804.09376

\bibitem[{{Macri} {et~al.}(2015){Macri}, {Ngeow}, {Kanbur}, {Mahzooni}, \&
  {Smitka}}]{Macri:2015}
{Macri}, L.~M., {Ngeow}, C.-C., {Kanbur}, S.~M., {Mahzooni}, S., \& {Smitka},
  M.~T. 2015, \aj, 149, 117

\bibitem[{{Madore}(1975)}]{Madore1975}
{Madore}, B.~F. 1975, \apjs, 29, 219

\bibitem[{{Madore}(1982)}]{madore82}
---. 1982, \apj, 253, 575

\bibitem[{{Michalik} \& {Lindegren}(2016)}]{Michalik:2016}
{Michalik}, D., \& {Lindegren}, L. 2016, \aap, 586, A26

\bibitem[{{Moffett} \& {Barnes}(1984)}]{Moffett+1984}
{Moffett}, T.~J., \& {Barnes}, III, T.~G. 1984, \apjs, 55, 389

\bibitem[{{Monson} \& {Pierce}(2011)}]{Monson+2011}
{Monson}, A.~J., \& {Pierce}, M.~J. 2011, \apjs, 193, 12

\bibitem[{{Pel}(1976)}]{Pel1976}
{Pel}, J.~W. 1976, \aaps, 24, 413

\bibitem[{{Persson} {et~al.}(2004){Persson}, {Madore}, {Krzemi{\'n}ski},
  {Freedman}, {Roth}, \& {Murphy}}]{persson04}
{Persson}, S.~E., {Madore}, B.~F., {Krzemi{\'n}ski}, W., {et~al.} 2004, \aj,
  128, 2239

\bibitem[{{Pietrzy{\'n}ski} {et~al.}(2013){Pietrzy{\'n}ski}, {Graczyk},
  {Gieren}, {Thompson}, {Pilecki}, {Udalski}, {Soszy{\'n}ski}, {Koz{\l}owski},
  {Konorski}, {Suchomska}, {Bono}, {Moroni}, {Villanova}, {Nardetto},
  {Bresolin}, {Kudritzki}, {Storm}, {Gallenne}, {Smolec}, {Minniti}, {Kubiak},
  {Szyma{\'n}ski}, {Poleski}, {Wyrzykowski}, {Ulaczyk}, {Pietrukowicz},
  {G{\'o}rski}, \& {Karczmarek}}]{Pietrzynski:2013}
{Pietrzy{\'n}ski}, G., {Graczyk}, D., {Gieren}, W., {et~al.} 2013, \nat, 495,
  76

\bibitem[{{Planck Collaboration} {et~al.}(2016){Planck Collaboration},
  {Aghanim}, {Ashdown}, {Aumont}, {Baccigalupi}, {Ballardini}, {Banday},
  {Barreiro}, {Bartolo}, {Basak}, {Battye}, {Benabed}, {Bernard}, {Bersanelli},
  {Bielewicz}, {Bock}, {Bonaldi}, {Bonavera}, {Bond}, {Borrill}, {Bouchet},
  {Boulanger}, {Bucher}, {Burigana}, {Butler}, {Calabrese}, {Cardoso},
  {Carron}, {Challinor}, {Chiang}, {Colombo}, {Combet}, {Comis}, {Coulais},
  {Crill}, {Curto}, {Cuttaia}, {Davis}, {de Bernardis}, {de Rosa}, {de Zotti},
  {Delabrouille}, {Delouis}, {Di Valentino}, {Dickinson}, {Diego}, {Dor{\'e}},
  {Douspis}, {Ducout}, {Dupac}, {Efstathiou}, {Elsner}, {En{\ss}lin},
  {Eriksen}, {Falgarone}, {Fantaye}, {Finelli}, {Forastieri}, {Frailis},
  {Fraisse}, {Franceschi}, {Frolov}, {Galeotta}, {Galli}, {Ganga},
  {G{\'e}nova-Santos}, {Gerbino}, {Ghosh}, {Gonz{\'a}lez-Nuevo}, {G{\'o}rski},
  {Gratton}, {Gruppuso}, {Gudmundsson}, {Hansen}, {Helou},
  {Henrot-Versill{\'e}}, {Herranz}, {Hivon}, {Huang}, {Ili{\'c}}, {Jaffe},
  {Jones}, {Keih{\"a}nen}, {Keskitalo}, {Kisner}, {Knox}, {Krachmalnicoff},
  {Kunz}, {Kurki-Suonio}, {Lagache}, {Lamarre}, {Langer}, {Lasenby},
  {Lattanzi}, {Lawrence}, {Le Jeune}, {Leahy}, {Levrier}, {Liguori}, {Lilje},
  {L{\'o}pez-Caniego}, {Ma}, {Mac{\'{\i}}as-P{\'e}rez}, {Maggio}, {Mangilli},
  {Maris}, {Martin}, {Mart{\'{\i}}nez-Gonz{\'a}lez}, {Matarrese}, {Mauri},
  {McEwen}, {Meinhold}, {Melchiorri}, {Mennella}, {Migliaccio},
  {Miville-Desch{\^e}nes}, {Molinari}, {Moneti}, {Montier}, {Morgante}, {Moss},
  {Mottet}, {Naselsky}, {Natoli}, {Oxborrow}, {Pagano}, {Paoletti},
  {Partridge}, {Patanchon}, {Patrizii}, {Perdereau}, {Perotto}, {Pettorino},
  {Piacentini}, {Plaszczynski}, {Polastri}, {Polenta}, {Puget}, {Rachen},
  {Racine}, {Reinecke}, {Remazeilles}, {Renzi}, {Rocha}, {Rossetti}, {Roudier},
  {Rubi{\~n}o-Mart{\'{\i}}n}, {Ruiz-Granados}, {Salvati}, {Sandri},
  {Savelainen}, {Scott}, {Sirri}, {Sunyaev}, {Suur-Uski}, {Tauber}, {Tenti},
  {Toffolatti}, {Tomasi}, {Tristram}, {Trombetti}, {Valiviita}, {Van Tent},
  {Vibert}, {Vielva}, {Villa}, {Vittorio}, {Wandelt}, {Watson}, {Wehus},
  {White}, {Zacchei}, \& {Zonca}}]{Planck-Collaboration:2016}
{Planck Collaboration}, {Aghanim}, N., {Ashdown}, M., {et~al.} 2016, \aap, 596,
  A107

\bibitem[{{Pojmanski}(1997)}]{Pojmanski1997}
{Pojmanski}, G. 1997, \actaa, 47, 467

\bibitem[{{Prusti}(2012)}]{prusti12}
{Prusti}, T. 2012, Astronomische Nachrichten, 333, 453

\bibitem[{{Riess}(2011)}]{riess11c}
{Riess}, A.~G. 2011, {An Independent Determination of WFC3-IR Zeropoints and
  Count Rate Non-Linearity from 2MASS Asterisms}, Tech. rep.

\bibitem[{{Riess} {et~al.}(2014){Riess}, {Casertano}, {Anderson}, {MacKenty},
  \& {Filippenko}}]{riess14}
{Riess}, A.~G., {Casertano}, S., {Anderson}, J., {MacKenty}, J., \&
  {Filippenko}, A.~V. 2014, \apj, 785, 161

\bibitem[{{Riess} {et~al.}(2016){Riess}, {Macri}, {Hoffmann}, {Scolnic},
  {Casertano}, {Filippenko}, {Tucker}, {Reid}, {Jones}, {Silverman},
  {Chornock}, {Challis}, {Yuan}, {Brown}, \& {Foley}}]{Riess:2016}
{Riess}, A.~G., {Macri}, L.~M., {Hoffmann}, S.~L., {et~al.} 2016, \apj, 826, 56

\bibitem[{{Riess} {et~al.}(2018){Riess}, {Casertano}, {Yuan}, {Macri},
  {Anderson}, {MacKenty}, {Bowers}, {Clubb}, {Filippenko}, {Jones}, \&
  {Tucker}}]{Riess:2018}
{Riess}, A.~G., {Casertano}, S., {Yuan}, W., {et~al.} 2018, \apj, 855, 136

\bibitem[{{Sahu} {et~al.}(2015){Sahu}, {Gosmeyer}, \& {Baggett}}]{Sahu:2015}
{Sahu}, K., {Gosmeyer}, C.~M., \& {Baggett}, S. 2015, {WFC3/UVIS Shutter
  Characterization}, Tech. rep.

\bibitem[{{Schechter} {et~al.}(1992){Schechter}, {Avruch}, {Caldwell}, \&
  {Keane}}]{Schechter+1992}
{Schechter}, P.~L., {Avruch}, I.~M., {Caldwell}, J.~A.~R., \& {Keane}, M.~J.
  1992, \aj, 104, 1930

\bibitem[{{Schmidt} {et~al.}(1995){Schmidt}, {Chab}, \&
  {Reiswig}}]{Schmidt+1995}
{Schmidt}, E.~G., {Chab}, J.~R., \& {Reiswig}, D.~E. 1995, \aj, 109, 1239

\bibitem[{{Shappee} {et~al.}(2014){Shappee}, {Prieto}, {Grupe}, {Kochanek},
  {Stanek}, {De Rosa}, {Mathur}, {Zu}, {Peterson}, {Pogge}, {Komossa}, {Im},
  {Jencson}, {Holoien}, {Basu}, {Beacom}, {Szczygie{\l}}, {Brimacombe},
  {Adams}, {Campillay}, {Choi}, {Contreras}, {Dietrich}, {Dubberley},
  {Elphick}, {Foale}, {Giustini}, {Gonzalez}, {Hawkins}, {Howell}, {Hsiao},
  {Koss}, {Leighly}, {Morrell}, {Mudd}, {Mullins}, {Nugent}, {Parrent},
  {Phillips}, {Pojmanski}, {Rosing}, {Ross}, {Sand}, {Terndrup}, {Valenti},
  {Walker}, \& {Yoon}}]{2014ApJ...788...48S}
{Shappee}, B.~J., {Prieto}, J.~L., {Grupe}, D., {et~al.} 2014, \apj, 788, 48

\bibitem[{{Stassun} \& {Torres}(2016)}]{Stassun:2016}
{Stassun}, K.~G., \& {Torres}, G. 2016, \aj, 152, 180

\bibitem[{{Szabados}(1980)}]{Szabados1980}
{Szabados}, L. 1980, Commmunications of the Konkoly Observatory Hungary, 76, 1

\bibitem[{{Szabados}(1981)}]{Szabados1981}
---. 1981, Commmunications of the Konkoly Observatory Hungary, 77, 1

\bibitem[{{Szabados}(1991)}]{Szabados1991}
---. 1991, Commmunications of the Konkoly Observatory Hungary, 96, 123

\bibitem[{{van Leeuwen} {et~al.}(2007){van Leeuwen}, {Feast}, {Whitelock}, \&
  {Laney}}]{vanleeuwen07}
{van Leeuwen}, F., {Feast}, M.~W., {Whitelock}, P.~A., \& {Laney}, C.~D. 2007,
  \mnras, 379, 723

\bibitem[{{Walraven} {et~al.}(1964){Walraven}, {Tinbergen}, \&
  {Walraven}}]{Walraven+1964}
{Walraven}, J.~H., {Tinbergen}, J., \& {Walraven}, T. 1964, \bain, 17, 520

\bibitem[{{Welch}(1986)}]{Welch1986}
{Welch}, D.~L. 1986, PhD thesis, UNIVERSITY OF TORONTO (CANADA).

\bibitem[{{Welch} {et~al.}(1984){Welch}, {Wieland}, {McAlary}, {McGonegal},
  {Madore}, {McLaren}, \& {Neugebauer}}]{Welch+1984}
{Welch}, D.~L., {Wieland}, F., {McAlary}, C.~W., {et~al.} 1984, \apjs, 54, 547

\bibitem[{{Zinn} {et~al.}(2018){Zinn}, {Pinsonneault}, {Huber}, \&
  {Stello}}]{Zinn:2018}
{Zinn}, J.~C., {Pinsonneault}, M.~H., {Huber}, D., \& {Stello}, D. 2018, ArXiv
  e-prints, arXiv:1805.02650

\end{thebibliography}
\clearpage

\begin{figure}[ht]
\vspace*{150mm}
\figurenum{1}
\includegraphics{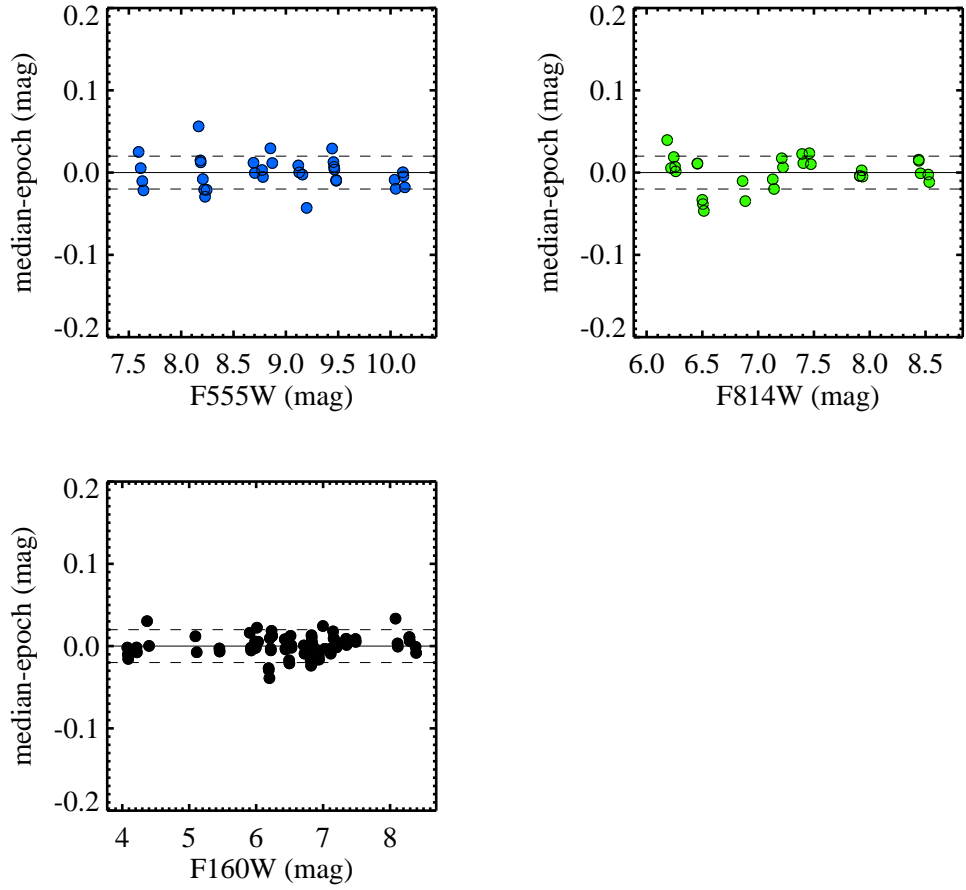}
\caption{\label{fg:outlier}  Variations from individual epochs of Cepheid photometry, phase-corrected to the epoch of the mean intensity.  For Cepheids with three or more epochs we compare the individual epoch values to the median of the Cepheid set.  The variations are caused by errors in photometry and errors in the phase corrections, and they have a mean dispersion in each filter of $<0.02$ mag.}
\end{figure}

\begin{figure}[ht]
\vspace*{150mm}
\figurenum{2}
\includegraphics{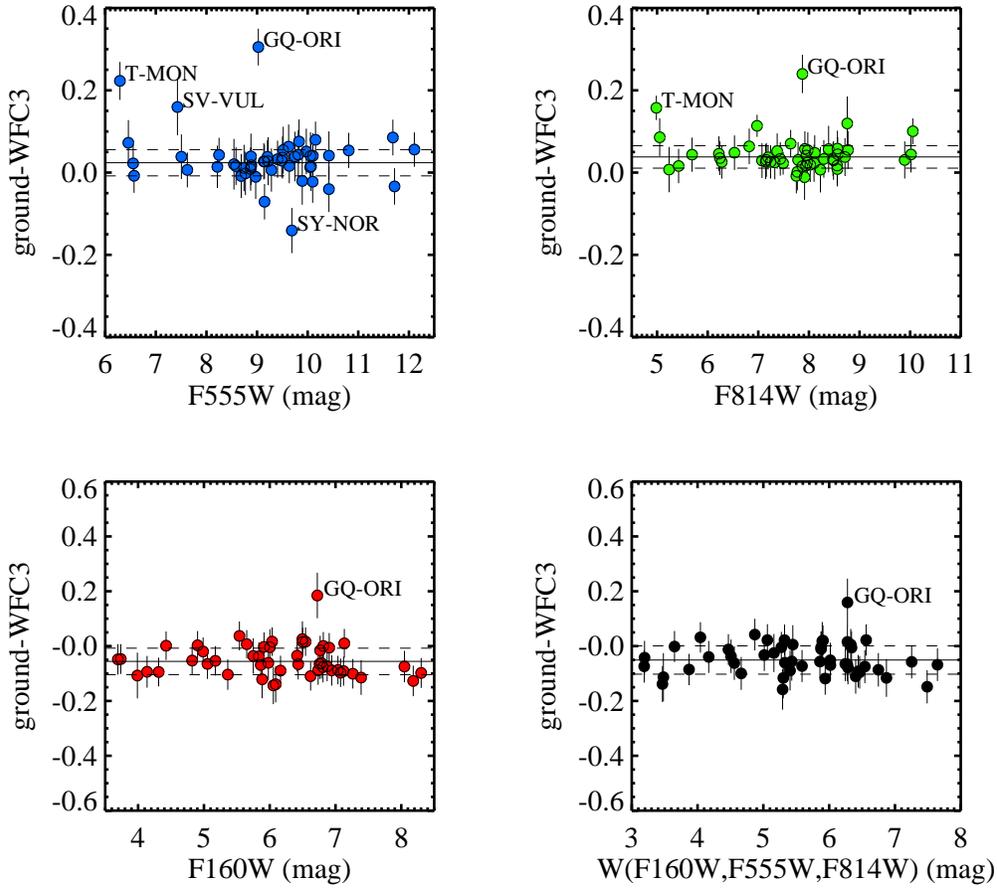}
\caption{\label{fg:outlier}  Comparison of Cepheid mean magnitudes in three {\it HST} WFC3 bands for observations obtained with {\it HST} and from the ground (transformed to the {\it HST} system). }
\end{figure}

\begin{figure}[ht]
\vspace*{150mm}
\figurenum{3}
\includegraphics{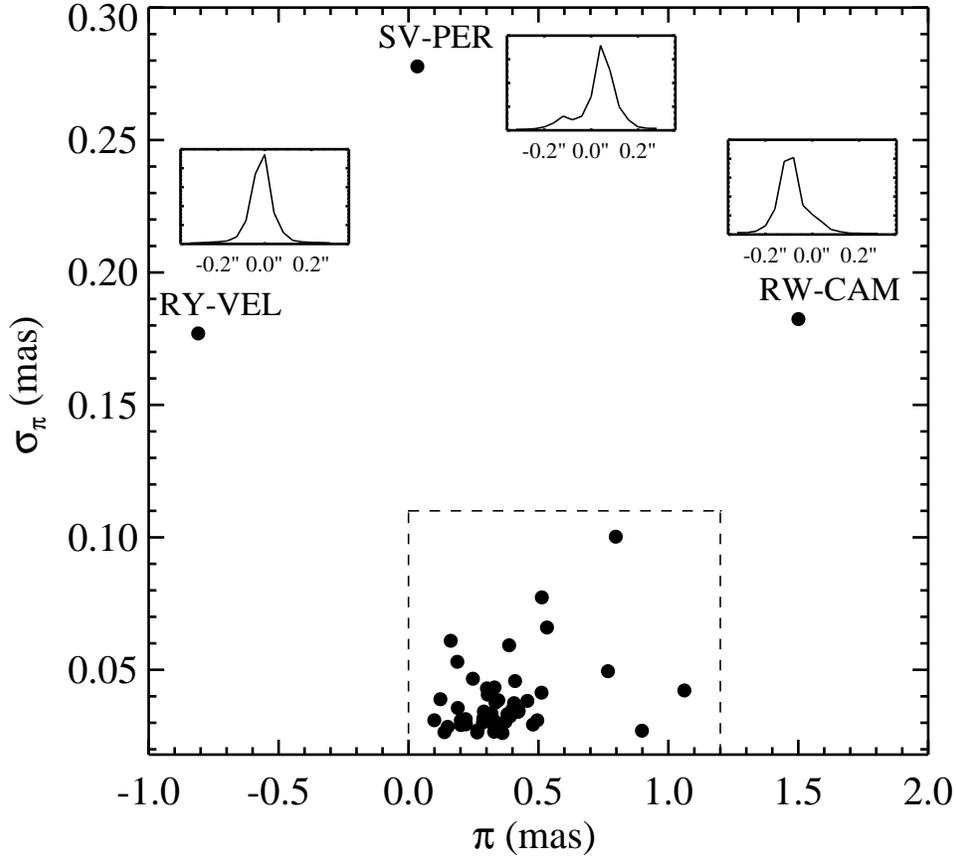}
\caption{\label{fg:outlier}  For Gaia DR2, reported values of $\pi$ and $\sigma_{\pi}$ for the sample of 50 Milky Way Cepheids with {\it HST} WFC3 system photometry.  The Gaia team reports DR2 has ``A small proportion of sources with corrupted parallaxes indicated by the occurence of apparently very significant large positive or negative values."  We identify three Cepheids whose parallaxes are likely corrupted as they appear far from the rest in this space (SV-Per, RW-Cam, and RY-Vel). For SV-Per and RW-Cam, our WFC3 spatial scans (insets) reveal a close companion within $0.2''$ of the {\it HST} line-spread function, which is the likely source of the corruption. These 3 are excluded from further analysis.}
\end{figure}

\begin{figure}[ht]
\vspace*{150mm}
\figurenum{4}
\includegraphics{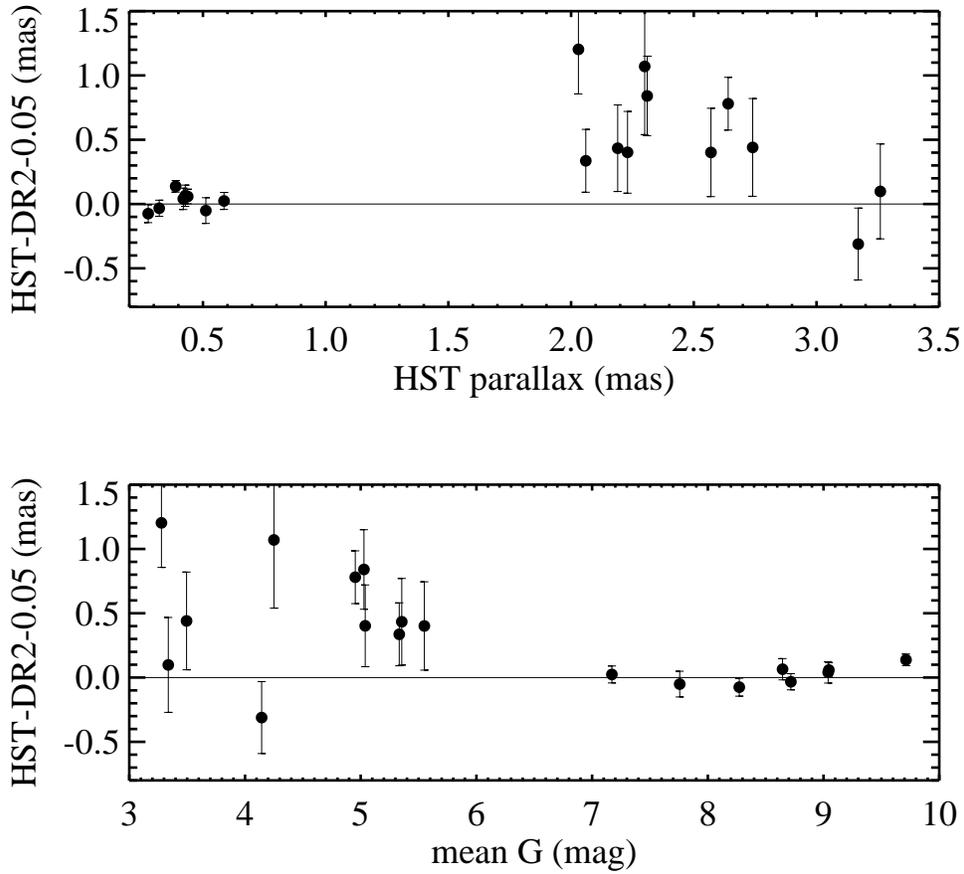}
\caption{\label{fg:outlier}  Comparison of 18 parallax measurements for Milky Way Cepheids measured with Gaia DR2 and with {\it HST} using the FGS \citep{benedict07} for $G<6$ or WFC3 spatial scanning for $G>6$ \citep{Riess:2018}.  Two Cepheids (Y-Sgr and Delta-Cep) were excluded because their Gaia DR2 values were negative, and one ($\ell$~Car) was extremely large, indicating they are corrupted.  The agreement is good for $G>6$ mag (7 of 8 within 1 $\sigma$) but poor at $G<6$ mag (1 of 10 within 1 $\sigma$), indicating that at $G<6$ mag, where the Gaia detectors saturate, the DR2 parallaxes become unreliable.  }
\end{figure}

\begin{figure}[ht]
\vspace*{150mm}
\figurenum{5}
\includegraphics{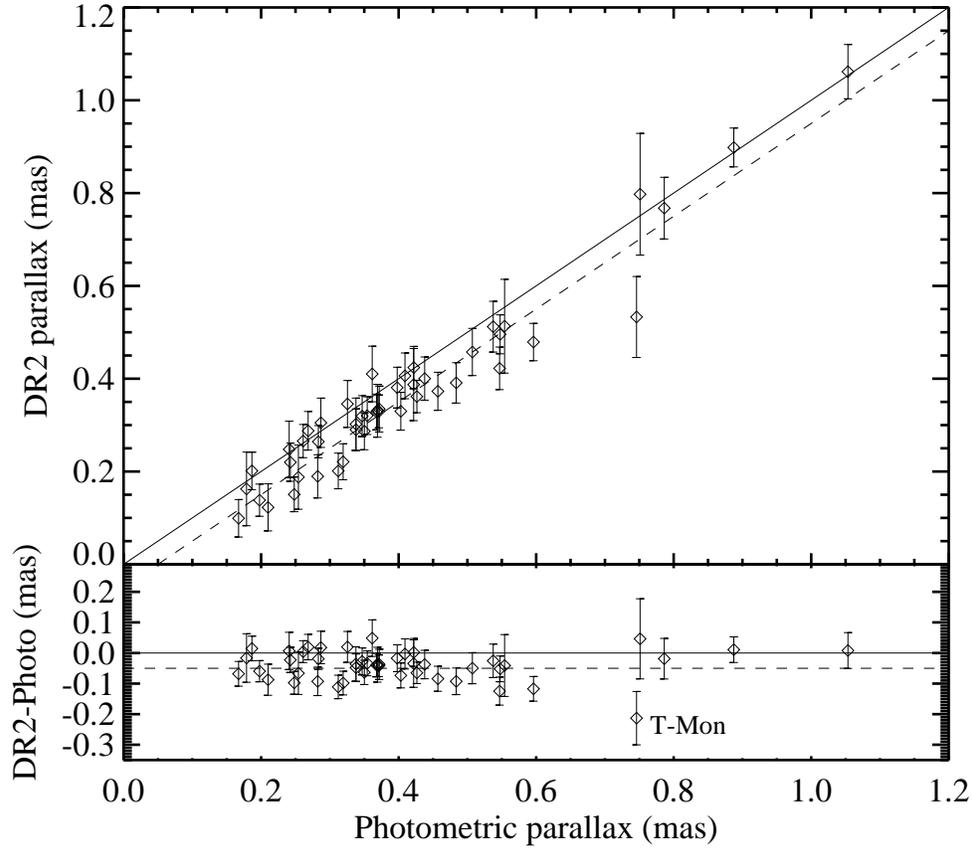}
\caption{\label{fg:outlier}  Comparison of 46 Milky Way Cepheid parallaxes provided in Gaia DR2 and predicted photometrically using the {\it HST} WFC3-based photometry in Table 1, the Cepheid periods, and the {\PL} parameters given by R16.  A zeropoint offset, as indicated (dashed), is readily apparent with otherwise good agreement.}
\end{figure}

\begin{figure}[ht]
\vspace*{150mm}
\figurenum{6}
\includegraphics{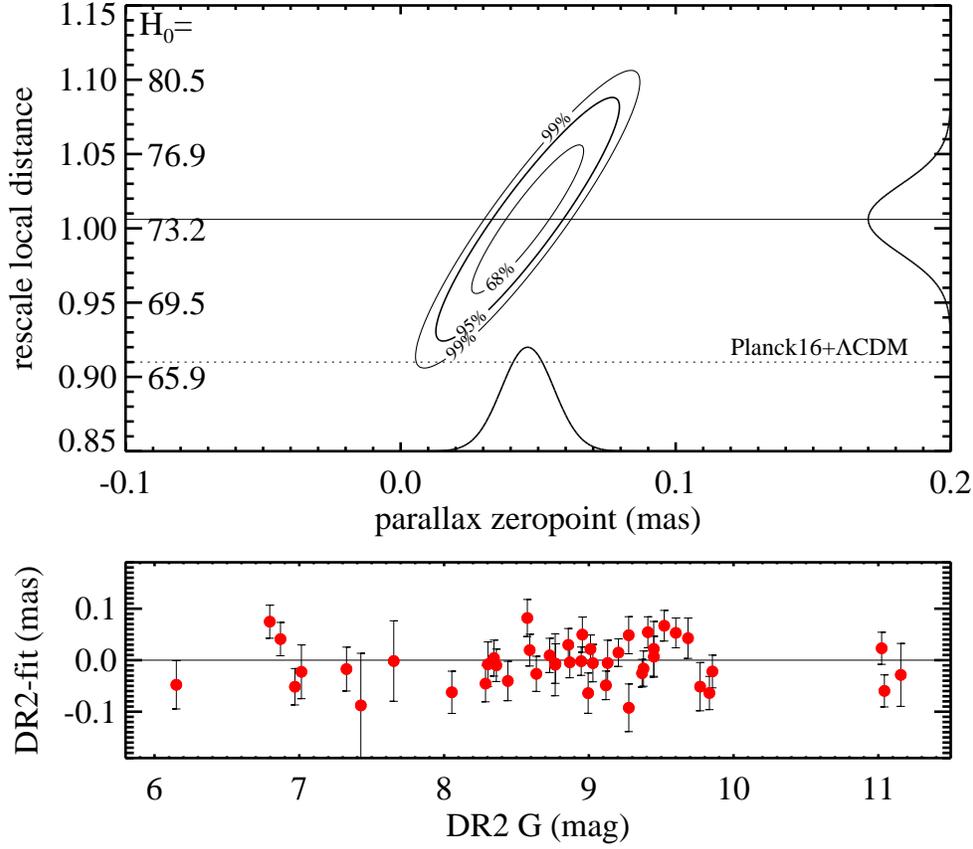}
\caption{\label{fg:outlier} For the {\it HST} sample of 50 Milky Way Cepheids, a sample with long periods, low extinction, and homogeneous photometry, we determined the best match between the measured Gaia DR2 parallaxes and those predicted {\it photometrically} from their photometry, periods, and the SHOES distance ladder of \cite{Riess:2016}. We allow two free parameters to account for the parallax zeropoint offset, $zp$, and a rescaling of the distance ladder, $\alpha$.  We find a significant zeropoint offset of \facezp and a rescaling of the SHOES distance ladder of \facemult.  The rescaling parameter is inconsistent at the \faceplanck $\sigma$ confidence level (99.6\%) with the value needed to match Planck + $\Lambda$CDM \citep{Planck-Collaboration:2016}. The lower panel shows the residuals from the best fit.}
\end{figure}

\begin{figure}[ht]
\vspace*{150mm}
\figurenum{7}
\includegraphics{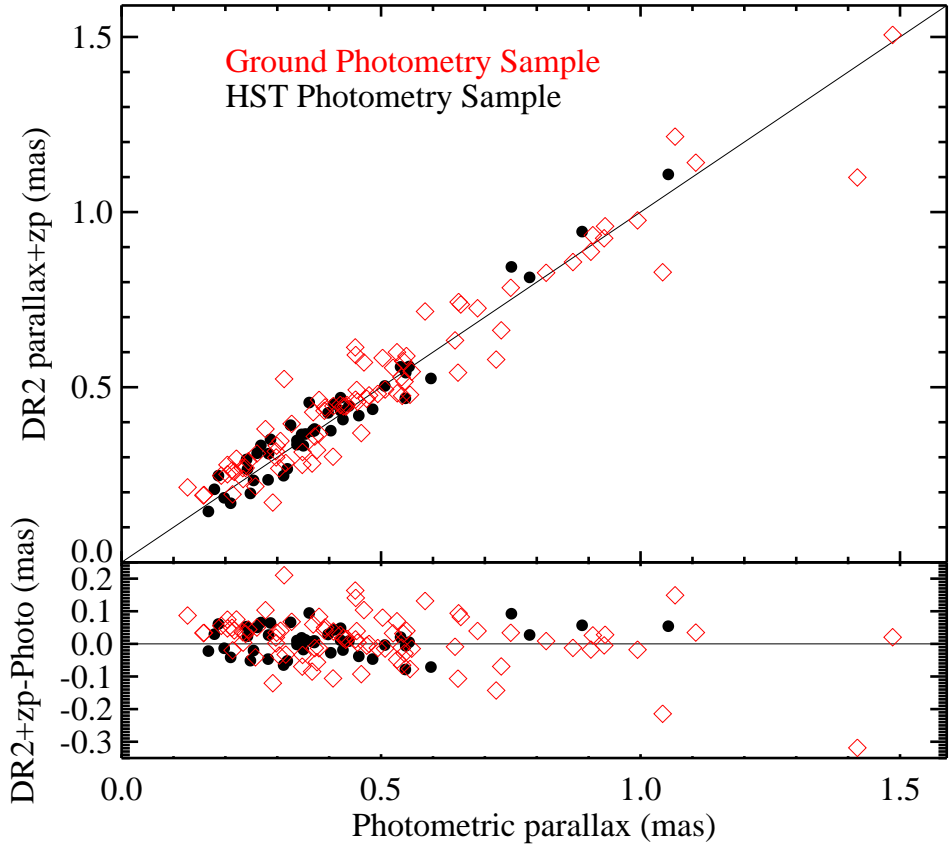}
\caption{\label{fg:outlier} Comparison of measured and photometrically predicted parallaxes for different Cepheid samples.  The {\it HST} sample of 50 Cepheids presented here was selected to have $P> 8$ days, $A_H < 0.4 $ mag, $V > 6$ mag, and expected distances of $D < 6$ kpc. It is 70\% complete (by random selection of the {\it HST} schedule) and has a dispersion of 43 $\mu$as, comparable to expectations.  A nonoverlapping sample of 86 Cepheids with photometry compiled from a single source for each ground-based system (see text) shows much greater dispersion, 99 $\mu$as or 68 $\mu$as after discarding the two most deviant (or 60 $\mu$as after discarding the four most deviant), far more dispersion than the DR2 errors can explain.  The text discusses reasons why such samples may be unreliable.}
\end{figure}

\begin{figure}[ht]
\vspace*{150mm}
\figurenum{8}
\includegraphics{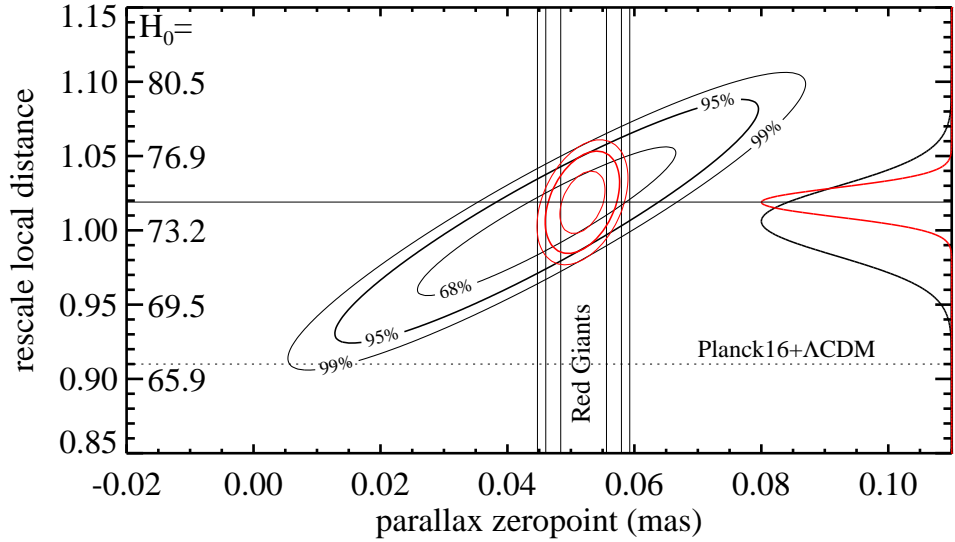}
\caption{\label{fg:outlier} Same as Figure 6 except now including a constraint of -53 $\pm 2.6$ $\mu$as on the parallax zeropoint offset calculated from 3475 Red Giants with Kepler-based asteroseismic estimates of radii and parallaxes from \cite{Zinn:2018}.  The constraint is intended only to illustrate the reduction in uncertainty in the distance scale that is possible with independent knowledge of the offset term.}
\end{figure}

\end{document}